\documentclass[journal]{IEEEtran}

\usepackage{latexsym,amssymb,exscale,graphicx,amsfonts,amsmath, bm}
\usepackage{amsmath, bm}
\usepackage[utf8]{inputenc}
\usepackage{multirow,multicol}
\usepackage{epsfig} 
\usepackage{setspace}
\usepackage[table]{xcolor}
\usepackage{xcolor,colortbl}
\usepackage{lipsum}
\usepackage{microtype}
\usepackage{stfloats}
\usepackage{cite}
\usepackage{epstopdf}
\usepackage{etoolbox}
\usepackage{cite}
\usepackage{color,soul}
\usepackage{tabulary}

\ifCLASSINFOpdf

\else

\fi

\begin{document}


\title{Power System Disturbance Classification with Online Event-Driven Neuromorphic Computing}


\author{Kaveri~Mahapatra,~\IEEEmembership{Student~member,~IEEE,}
        Sen Lu, Abhronil~Sengupta,~\IEEEmembership{Member,~IEEE,}
        and~Nilanjan~Ray~Chaudhuri,~\IEEEmembership{Senior~member,~IEEE} }

\maketitle

\begin{abstract}

Accurate online classification of disturbance events in a transmission network is an important part of wide-area monitoring. Although many conventional machine learning techniques are very successful in classifying events, they rely on extracting information from PMU data at control centers and processing them through CPU/GPUs, which are highly inefficient in terms of energy consumption. To solve this challenge without compromising accuracy, this paper presents a novel methodology based on event-driven neuromorphic computing architecture for classification of power system disturbances. A Spiking Neural Network (SNN)-based computing framework is proposed, which exploits sparsity in disturbances and promotes local event-driven operation for unsupervised learning and inference from incoming data. Spatio-temporal information of PMU signals is first extracted and encoded into spike trains and classification is achieved with SNN-based supervised and unsupervised learning framework. In addition, benefits of deep spiking networks for complex multi-class event identification problem are presented by leveraging increasing dynamic neural sparse spiking events with network depth. Moreover, a QR decomposition-based selection technique is proposed to identify signals participating in the low rank subspace of multiple disturbance events. Performance of the proposed method is validated on data collected from a $16$-machine, $5$-area New England-New York system.

\end{abstract}


\IEEEpeerreviewmaketitle

\vspace{-12pt}
\section{Introduction}\label{sec:Intro}

\IEEEPARstart{A}DVENT of Phasor Measurement Units (PMUs) is leading to a new era of wide-area monitoring with large-scale utilization of machine learning (ML) and deep learning (DL) techniques. 
Proliferation of PMUs offer better observability of the modes of the system and helps building situational awareness. However, this also brings along challenges in handling and utilization of higher volume of time-series data. Typically, PMU measurements are transmitted at high speeds to digital computing systems (e.g. DSP, microprocessor or computers with CPU/GPU capabilities) in the control center, which requires high computational power and consumes lots of energy while processing the data for monitoring applications. One such application is the classification of large-scale events using PMU data. 

The conventional way to approach event classification problem involves two steps. The first step is to utilize many measurements and apply feature engineering to extract relevant features for events under consideration. Some examples are raw time-series, energy function approach \cite{brahma2017real,yadav2018real,tang2017dynamic}, ellipsoid characteristics~\cite{gajjar2014auto, dahal2014comprehensive, dahal2012preliminary, lavand2016mining}, frequency-domain details such as wavelets or shapelets ~\cite{aravind2016pmu, brahma2017real, kim2017wavelet, bhuiyan2018wpd, lavand2016mining, biswal2016signal }, principal component analysis~\cite{chen2014power, gajjar2014auto, rafferty2016real,niazazari2017disruptive}. Extraction of features involve computations for every new data window. In contrast, we propose a feature selection method to utilize a subset of raw signals directly into the spiking network framework so that the computations associated with feature extraction can be avoided. 

The second step is the utilization of ML/DL based classification algorithms. Classification algorithms proposed in literature are broadly based on either statistical tests, correlation tests or thresholds~\cite{okumus2018event, phillips2014distribution, yadav2018real, rafferty2016real,kim2017wavelet}, or different ML methods such as clustering~\cite{dahal2014comprehensive, guo2016online }, decision tree~\cite{nguyen2015smart,zhang2009real,yin2016initial }, support vector machines~\cite{niazazari2017disruptive, dahal2014evaluating, seethalekshmi2012classification, yin2016initial, biswal2016signal}, k-Nearest neighbor~\cite{aravind2016pmu, dahal2014evaluating, biswal2016supervisory}, artificial neural network (ANN)~\cite{tang2017dynamic, yin2016initial}, extreme learning machines~\cite{biswal2016supervisory}, auto-encoders\cite{niazazari2017disruptive}, and deep learning architectures \cite{DeepLPMU} among others. These techniques inherently require a large computational effort in processing every incoming time-series data window for training and inference on a traditional computing platform. Especially, computations performed on an ANN platform for learning and inference is extremely hardware expensive and cannot leverage the sparse nature of disturbance events in PMU measurements.

In contrast, in the human brain, information is processed in networks composed of spiking neurons, which accepts dynamic binary spiking inputs as a function of time and with only parts of the network being active during any operation. A Spiking Neural Network (SNN) driven by the notion of biological neurons with computing by means of sparse binary spike signals over time is thus more biologically plausible, and supports unsupervised learning via event-driven hardware operation, consuming much less power. These features make SNN a suitable candidate for online learning and inference from PMU data for disturbance monitoring and thereby, necessitates development of SNN architecture for enabling algorithm-hardware co-design and encourages implementation on hardware for low power event-driven operation. In this work, we present a maiden application of neuromorphic computing platform in power systems for an event classification problem using PMU data streams. A SNN architecture is proposed to extract spatio-temporal information from PMU measurements in an efficient way for classification of different power system events without sacrificing accuracy obtained with conventional ML algorithms.

This paper is organized into six sections. Section~\ref{sec:SNN_Motiv} discusses the main motivation for shifting from ANN to a spiking domain for representing PMU data. Section~\ref{sec:SpikeTrnRepDescrp} presents spike-based representation of a time series signal. Section~\ref{sec:SigSelFramework} introduces a feature selection approach to complement SNN classification framework to reduce the computational burden by enabling selection of candidate raw signals. Section~\ref{sec:SNNSec} discusses different types of SNN learning frameworks for event classification problem. Section~\ref{sec:Results} presents the experimental set up and discusses the results of performance comparison between ANN and SNN. Section~\ref{sec:DeepSNN} presents results with deeper spiking neural networks on event classification problem. Section~\ref{sec:Concl} concludes the paper.

\vspace{-12pt}
\section{Motivation}\label{sec:SNN_Motiv}

Disturbance event identification and characterization in smart grid using PMU data is a computationally expensive task for online operations. Availability of ML/DL resources on CPU/GPU at control centers helps in solving problems requiring training and inference using large PMU data sets. However, due to lack of event-driven operation and inability to exploit sparsity patterns in the data, the computational burden, energy consumption, and time required to train and deploy networks for this complex solution space is humongous. Addressing the high computational burden of machine learning hardware has been identified as one of the Grand Challenges in Computing for the next decade by the US government and Federal agencies \cite{NanoGrandCh, IBMCat}. For instance, training common large Deep Learning models can result in more than 626,000 pounds of carbon emissions - the equivalent of 5 times the lifetime emission of an average car \cite{strubell2019energy}. Therefore, power grid's secure operation also comes at a cost of significant increase in its carbon footprint, which counters the sustainability goals in today's world. This just shows the inefficiency of our current computational platforms (CPU/GPUs/cloud clusters) and how these can be improved with neuromorphic computing. The goal of this paper is to address this challenge for the smart grid community.

An under-explored fact here is that a power transmission system remains in ambient/quasi steady state operating condition most of the time and switches to transient state only during disturbance events for a brief period. Typical disturbance events such as line faults (voltage events), generator-load unbalancing (frequency events) due to generator outages or load shedding, and so on, occur infrequently over the span of the year. Table \ref{Table_NERC} shows two metrics from the North American Electric Reliability Corporation (NERC) report~\cite{NERC_SOR_2019} for years 2015-2018, which indicate the average probability of occurrence of transmission line outage events in a year is $\approx$ 0.237\% and also the number of occurrences of contingency events related to balancing resources/generation with demand following a disturbance ranges between $150 - 425$ per year.

\begin{table}[h]
\vspace{-12pt}
	\centering
	\footnotesize
	\renewcommand{\arraystretch}{0.9}
	\caption{ Findings from NERC State of Reliability Report~\cite{NERC_SOR_2019}}
	\vspace{-6pt}
	\label{Table_NERC}
	\begin{tabular}{c||c|c|c|c}
		\hline
		\bfseries Years & 2015 & 2016 & 2017 & 2018 \\ 
		\hline
		\bfseries  Transmission Unavailability (\%)   & 0.22 & 0.27 &  0.24 & 0.22  \\
		\hline
		\bfseries  Number of Disturbance Control & 350- & 375- & 150- & 175- \\
	    \bfseries   Standard Events (Range) & -400 & -425 & -200  & -225  \\
		\hline
	\end{tabular}
	\vspace{-7pt}
\end{table}

\begin{figure}[t!]
\vspace{-15pt}
	\centering
	\includegraphics[trim={0.65cm 0.2cm 0.2cm 0.3cm},clip,width=0.52\textwidth]{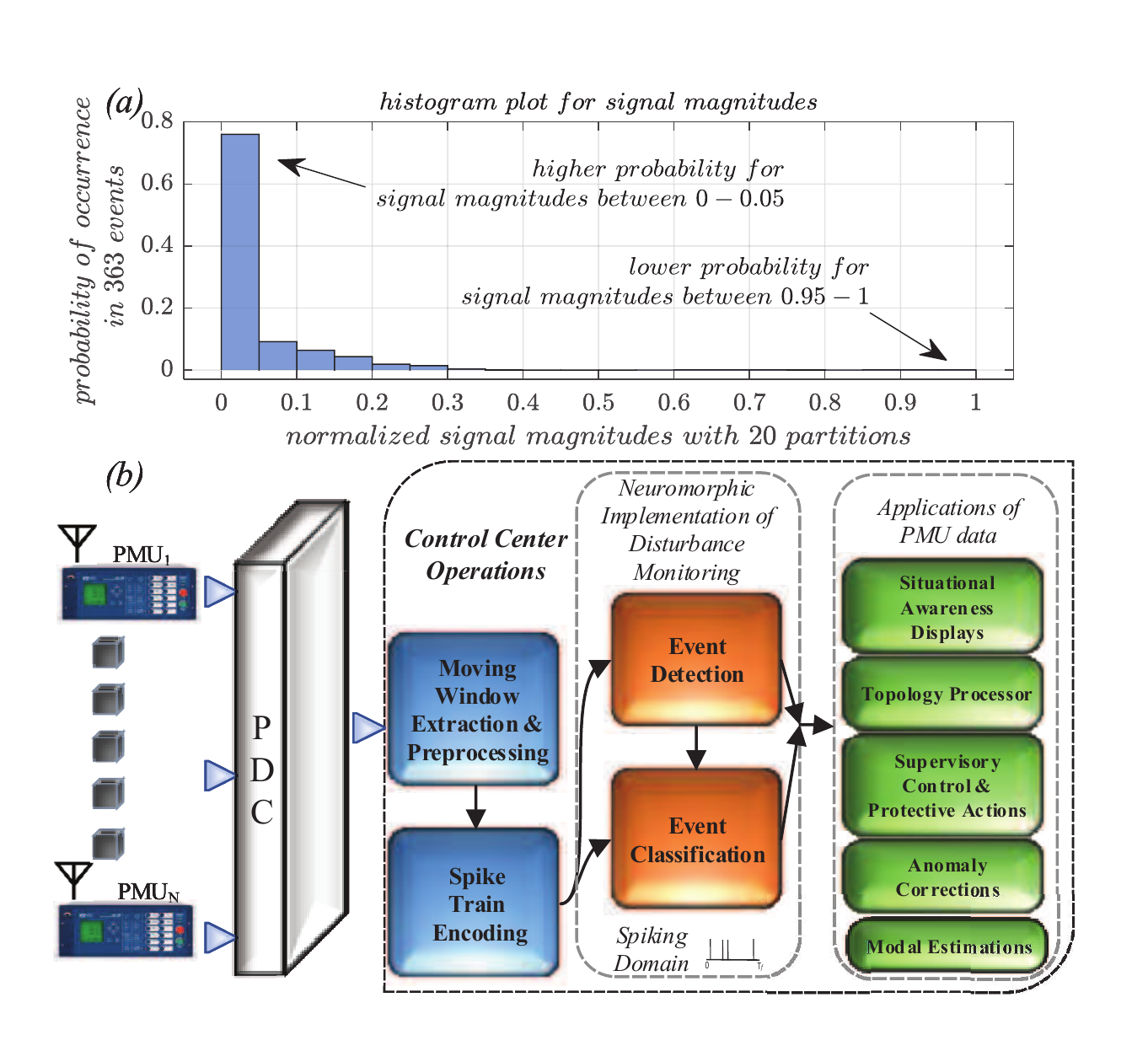}
\vspace{-15pt}
	\caption{(a) A histogram plot for a 15s measurement data window of 10 signals with disturbance starting at 5s over 363 disturbance events, where X-axis indicates 20 normalized magnitude ranges of magnitude deviations from ambient prefault values and Y-axis indicates the probability of any signal taking a magnitude in that range at any instant. The events are simulated in NE-NY system; (b) Proposed architecture for power system disturbance classification.}
	\label{fig:J5_CCArchFig}
\vspace{-15pt}
\end{figure}

In addition, based on a 15s window of simulated data from the New England-New York test system~\cite{pal2006robust}, a typical histogram of distribution of signal magnitudes for 363 events is shown in Fig.~\ref{fig:J5_CCArchFig}(a). Signal magnitude here refers to the deviation from nominal values  per unit magnitudes of phasor voltages ($|V|$), currents ($|I|$) or frequency ($f/60$) of various network variables in the system following a disturbance. 
Each of these windows contains 5s of pre-disturbance and 10s of post-disturbance data from each event. For more details on the data generation, please refer to Section VI-A. The characteristic of these deviations alters with post-disturbance operating condition. Under quasi-steady-state pre-disturbance (also called ambient) condition, the signal magnitudes are represented by zero mean gaussian noise. During disturbance, the fluctuations in these magnitudes are highest and they typically manifest electromechanical oscillations of smaller amplitude following the disturbance , which damp out if the system is stable. Therefore, the period of high magnitude fluctuations (0.95-1 times signal magnitude of 1 pu) from steady state value (of.1 p.u.) for most disturbances is short and thus such fluctuation values have a less probability of occurrence in the window right after disturbance.

This implies there is a higher probability of samples in the signals of the window to have a lower range of `magnitude deviation from the prefault values' that lies on the left side of histogram. Note that these magnitude deviations are caused by oscillations. Due to temporal sparsity of events, most of the oscillations occur only at the starting of a disturbance event and dies down after a small duration. In addition, due to spatial sparsity of PMUs, the signals which are closer to the disturbance location undergoes more significant deviations during disturbance compared to others which are farther (see 4.2.2 of \cite{PQBollen}). Moreover, spatial sparsity brings in dispersion in the observability of modes reflected in post-disturbance oscillation magnitudes. As a result, in the event window, the high magnitude samples are less frequent and can be characterized by a sparse distribution with close to 0 magnitude samples persisting mostly in the time horizon.

Interestingly, our brain computes by means of sparse, event-driven temporal signals and algorithms inspired by such computational principles can be an ideal fit here. Inspired by this property, we propose to leverage a disturbance event-driven neuromorphic computing platform, which can exploit sparsity in signal representation in spike domain and thus reduces the computational effort leading to energy savings.

SNN is a class of ANN-based computational model, which closely mimics our brain’s ability to naturally encode and process information. Its efficiency comes from the ``learning over time" feature as well as computation and communication in terms of spike in the operating model. The main advantages of SNN are as follows: (1) sparse signal processing, (2) event-driven hardware operation, and (3) lower energy consumption~\cite{LoihiMicroMikeDevies}. Recent developments show that SNN now has the potential to achieve similar accuracy as of other ML techniques with much less power consumption~\cite{sengupta2019goingdeeper}. Incoming data represented by a sparsely-distributed input spike train can significantly reduce energy consumption during periods of ambient condition of the grid.

Figure \ref{fig:J5_CCArchFig}(b) describes the proposed architecture for processing PMU signals at a control center in spiking domain. After receiving PMU data at phasor data concentrator (PDC), a moving window containing present and past samples of data is extracted. This data is first encoded into spike trains and then processed in SNNs for event detection and classification purposes. The main contributions of this work are as follows.

\begin{itemize}
\item A spike-based representation of PMU measurements and event-driven supervised and unsupervised power system event classification through a SNN architecture that can replace the CPU/GPU-based processing of state-of-the-art classification algorithms.
\item A signal selection technique to eliminate dependency on processing many signals without loosing any information and thus, reduce the computations during classification by operating directly on measurements instead of derived features, when the number of observed signals is large.
\end{itemize}
The following sections present these contributions.

\vspace{-12pt}
\section{Spike-Train Representation of a Time-Series Signal}\label{sec:SpikeTrnRepDescrp}

A spike train is a mode of transmitting information in the nervous system. This can be interpreted as a sequence of recorded times at which a neuron fires an action potential~\cite{Spike_Train_CMU}. The irregularity in arrival time of successive action potentials can be modelled as a random process. Under an independent spike generation hypothesis, the random process in the neocortex is assumed to follow a Poisson distribution. This is typically utilized in a process called ``rate coding''~\cite{orban2012neuronal}~to convert an input sample to a certain number of spikes while including an element of stochasticity.


The $i^{th}$ sample $y(i)$ of a discrete time-series measurement $[\bar y], i \in [1, N]$ is represented by first associating it with an input neuron $i$. With a homogeneous Poisson process assumption, the spike generation rate $\lambda_{i}$ of this neuron over an interval of spiking takes a fixed continuous analog value and is set to be proportional to the amplitude of the sample $y(i)$. In other words, the neuron $i$ is driven by a stimulus of $y(i)$ over the interval $[0,\texttt{T}]$ to generate spikes at a rate $\lambda_{i}=y(i)$. The spike train interval $\texttt{T}$ is subdivided into shorter intervals $\delta \texttt{t}$ such that $\texttt{t}=j\delta\texttt{t}$, where $j=0,1,2,...,\texttt{T}/\delta \texttt{t}$. The probability of spikes occurring during each $\delta \texttt{t}$ is set to $r \delta \texttt{t}$. For each $\delta \texttt{t}$, a uniformly distributed random number $\texttt{x}(\delta \texttt{t})$ is generated. This is followed by a spike generation as per the following rule.

 \vspace{-5pt}
\begin{equation}
\vspace{-4pt}
    \texttt{s}\left( {j\delta \texttt{t}} \right) = \left\{ {\begin{array}{*{20}c}
   {1,} & {\texttt{x}(j\delta \texttt{t}) \le r\delta \texttt{t}}  \\
   {0,} & {otherwise}  \\
\end{array}} \right.
 \Rightarrow     \texttt{s} = \sum\limits_{\texttt{T}_f } {\delta \left( {\texttt{t} - \texttt{t}_f } \right)} 
\end{equation}


\noindent where, $\texttt{s}$ is the spike train generated over the interval $[0,\texttt{T}]$ with spiking instances $\texttt{t}_f$ for sample $y(i)$. $\texttt{s}$ can also be represented as a set of Dirac-$\delta$ functions at time instants $\texttt{t}_f \in \texttt{T}_f$. These spike trains are given as input to the excitatory layer of SNN and has been demonstrated in Fig.~\ref{fig:UnSupSNNArch}. This type of binary encoding of stimulus requires a single bit for computing as well as for communicating a spike. With address-event representation (AER)-based communication protocol, sparse neural events can be communicated efficiently in event-driven neuromorphic hardware~\cite{LoihiMicroMikeDevies}. 

A power system follows a quasi-steady state trajectory during most of the time of operation and only switches to a transient state trajectory during a disturbance. The spike generation rate follows a steady-state value proportional to signal magnitude during the steady state operation and fluctuates in presence of disturbance samples. Assuming that under steady state conditions of the system, the spike rate is set to a low value, the distribution in spike occurrence will be sparse over the interval of stimulus presentation to the neurons. This leads to inactivity in the regions of SNN, which can be exploited on a neuromorphic hardware through ``power gating''~\cite{chen1998estimation} by shutting off portions of the circuit, which are not being triggered by spiking events. Only upon the onset of an event, sufficient spikes are generated to activate the network enabling event-driven operation, and eliminating computational effort. Thus, a spiking representation of signals and the algorithm-hardware co-design of SNN and implementation on neuromorphic hardware could potentially result in significantly low power consumption.


\vspace{-5pt}
\section{Signal Selection}\label{sec:SigSelFramework}

PMU data approximated by low dimensional feature set could potentially overcome the dependency on processing all PMU signals online. However, it comes at a cost of SVD computations every time a new sample arrives, and thus imposes a complexity of ${\rm \mathcal{O}}\left( {mN } \right)$ for online feature extraction from a $N\times m$ data matrix. To overcome this, a feature selection procedure based on economy QR decomposition~\cite{golub1996matrix} is proposed to identify PMU signals which participate in forming the low-rank subspace under multiple disturbance events that could be used as input to the SNN.

PMUs provide discrete synchronous time-series measurements of electrical quantities $y(t)$ such as voltage, current, power and frequency at any time $t$. The system response $Y$ ($N \times m$) containing $N$ samples for $t = [0, (N-1)Ts ]$ of $m$ measured signals can be expressed as a product of Vandermonde matrix $V$ and complex transformation matrix $F_o$.

\vspace{-8pt}
\begin{equation}\label{eq:yvff}
\vspace{-5pt}
Y^T = F_o V \Rightarrow Y=V^*{F_o}^*
\end{equation}

The rectangular matrix $Y$ can be decomposed into two factors using economy QR decomposition~\cite{golub1996matrix}, which rearranges signals present in $Y$ in decreasing order of priority for obtaining a linearly independent set of signals in the columns of $\mathord{\buildrel{\lower3pt\hbox{$\scriptscriptstyle\leftharpoonup$}} \over Y}$.

\vspace{-8pt}
\begin{equation}\label{eq:QRdec}
\vspace{-5pt}
YP = \mathord{\buildrel{\lower3pt\hbox{$\scriptscriptstyle\leftharpoonup$}} 
\over Y}  = QR \Rightarrow Q^T \mathord{\buildrel{\lower3pt\hbox{$\scriptscriptstyle\leftharpoonup$}} 
\over Y}  = R 
\end{equation}

where, $Q$ represents an orthonormal basis for subspace of ${\mathord{\buildrel{\lower3pt\hbox{$\scriptscriptstyle\leftharpoonup$}} 
\over Y} }$ and $R$ represents an upper triangular matrix $r_{k,l}={\bar q_k^T }\bar y_l$, $P$ represents permutation matrix, which permutes the columns of $Y$ to get nonincreasing diagonal entries in $R$. When all the signals present in $Y$ are linearly independent, the rank of subspace $U$ of $Y$ is $m$ implying at least $m$ distinct modes ($v_i$) of $V$ are contributing in forming $U$. Since, $Y$ is low rank, the rank of $U$ becomes $n<m<<N$. Since $Q \subseteq {\cal{R}}(Y)$, where ${\cal{R}}$ indicates the column space, it can be written as a mapping from $V^*$ as follows.

\vspace{-10pt}
\begin{equation}\label{eq:QGV}
\vspace{-10pt}
Q=V^*{G_o}^*\Rightarrow Q^T={G_o}V
\end{equation}
\vspace{-8pt}
\begin{equation}\label{eq:Gdagger}
\vspace{-5pt}
\begin{array}{l}
 Q^T Q = I = GVV^* G^*  = G\hat VG^*  \\ 
  \Rightarrow G\hat V = \left(  {G^* } \right)^{\dagger} =\left(  {G^{\dagger} } \right)^{*} \\ 
 \end{array}
\end{equation}

Now substituting expression for ${G^{\dagger}}^*$ in \eqref{eq:QRdec} for $R$ leads to
\vspace{-5pt}
\begin{equation}\label{eq:RinFG}
\vspace{-5pt}
R = Q^T \mathord{\buildrel{\lower3pt\hbox{$\scriptscriptstyle\leftharpoonup$}} 
\over Y}  = GVV^* F^*  = G\hat VF^*  = \left( {G^{\dagger} } \right)^{*}F^* 
\end{equation}
This implies $R$ is the projection of $F^*$ onto $G^*$, which are both mapping complex exponential time series in Vandermonde matrix to the signal subspace of $Y$. Each entry $r_{k,l}$ of $R$ is equivalent to ${\bar g_k^T }\bar f_l$, which is the projected component of $\bar f_l$ in the direction of ${\bar g_k}$ of $Q$. 
\vspace{-5pt}
\begin{equation}\label{eq:rkl-1}
\vspace{-5pt}
\begin{array}{l}
\bar g_k^T \bar f_l  = 0,\forall l,k = l + 1:m \\*
\bar g_1^T \bar f_1  \ge \bar g_2^T \bar f_2  \ge ...\bar g_n^T \bar f_n  \ge ... \ge \bar g_{m - 1}^T \bar f_{m - 1}  \ge \bar g_m^T \bar f_m  \\*
\bar g_k^T \bar f_k  = 0,n < k \le m 
 \end{array}
\end{equation}

Note that the projected component in $r_{k,l}$ are proportional to the $F^*$ matrix. A total of $n$ independent column vectors of $F^*$ corresponding to top $n$ diagonal entries of $R$ implies $n$ linearly independent signals forming the subspace $U$ of all $m$ signals in $Y$. This set forms a minimal group, which at the least ensures observability of high energy modes responsible for oscillations observed through $Y$. 

Since, the objective of this work is to detect transient disturbances with low frequencies of oscillation (persisting over longer duration) carrying more energy and the number of low frequency modes present in the system is $\Tilde{n}<n $,--a very low dimensional subspace $\hat Q$ (tall matrix with $\Tilde{n}$ orthonormal columns) need to be extracted from $Y$. In reality, both $F^*$ and $V^*$ are very sensitive to operating condition and disturbance as well. By applying the above analysis, we can obtain a linearly independent minimal group of $\Tilde{n}$ signals for each disturbance scenario. 

To select signals, a metric is proposed to find the importance of each signal among a group of signals based on its contribution in low rank subspace formation under multiple disturbance scenarios. Given a set of training examples of multiple disturbance events obtained by a contingency analysis on the system, we propose the following heuristic to identify a group of signals of priority and to represent the lower dimensional signal subspace.

\begin{itemize}
\item \textbf{Input} Disturbance event data set $Y_h$ $\in \mathbb{R}^{N\times m}$, with $h \in [1:H_{train}]$, where $H_{train}$ is the number of training examples obtained under 3 types of disturbance events; Bus faults (Class-A), generator outage (Class-B), load trip (Class-C); \textbf{Output}: Important signal set $S$; \textbf{Parameters}: Threshold $\nu_1$ for selecting important signals in each $Y_h$;
\item \textbf{Initialization}
\begin{itemize}
	\item Set initial counter $\textit{CC}(j)$ for each signal $j=1:m$. 
	\item Set thresholds $\nu_1 = 0.1$.
	\item Set $h = 1$, set of all signals $M=[1:m]$, 
\end{itemize}
\item \textbf{while} $h \le H_{train}$
\begin{enumerate}
	\item Apply economy QR decomposition on $Y_h$ to obtain permutation matrix $P$, orthonormal matrix $Q$, $R$. 
	\item Find the permutation vector $p$ from $P$ containing arrangement of signals in $Y$ used for economy QR decomposition.
	\item Find set $\hat S = \left\{ {s|s \in p,r_{s,s}  \ge \nu_1 r_{1,1} } \right\}$.
	\item Update counter $\textit{CC}(j)= \textit{CC}(j)+1$ for all $j \in \hat S$.
	\item Increment $h$ by 1 and go to step (1).
\end{enumerate}
\item Find importance score $Sc$ for each signal $j$ as follows.
$Sc(j)= CC(j)/H_{train}$.
\item Select the top $n$ signals having highest scores in $Sc$ and save those in set $S$.
\item \textbf{Output} Set $S$ containing signal indices important for multiple disturbance event data $Y_h$ for $h=1:H_{train}$.
\end{itemize}

Given $m$ number of signals for any event in $Y_h$, signals in $\hat S_h$ forming the low-rank subspace $U_h$ is important for the event $h$. Cardinality ($\tilde n$) of set $\hat S_h$ is an estimate of the number of high energy modes $\lambda _is$  present in $m$ signals. Given $H_{train}$ events, a score $Sc(j)$ for each of signal indicates how many such events, signal $j$ has participated in forming their low-rank subspace. Thus, signals with higher scores become potential candidates for representing multiple disturbance events and thus, can faithfully ensure the information contained in signals not selected during this process is also preserved.


\vspace{-13pt}
\section{Spiking Neural Network}\label{sec:SNNSec}
\vspace{-4pt}
\subsection{Learning using Unsupervised Spiking Neural Network}\label{Sec:SNN_UnSup_Descrip}
\begin{figure}[t!]
	\centering
	\includegraphics[trim={0.3cm 0.2cm 0.2cm 0.3cm},clip,width=0.5\textwidth]{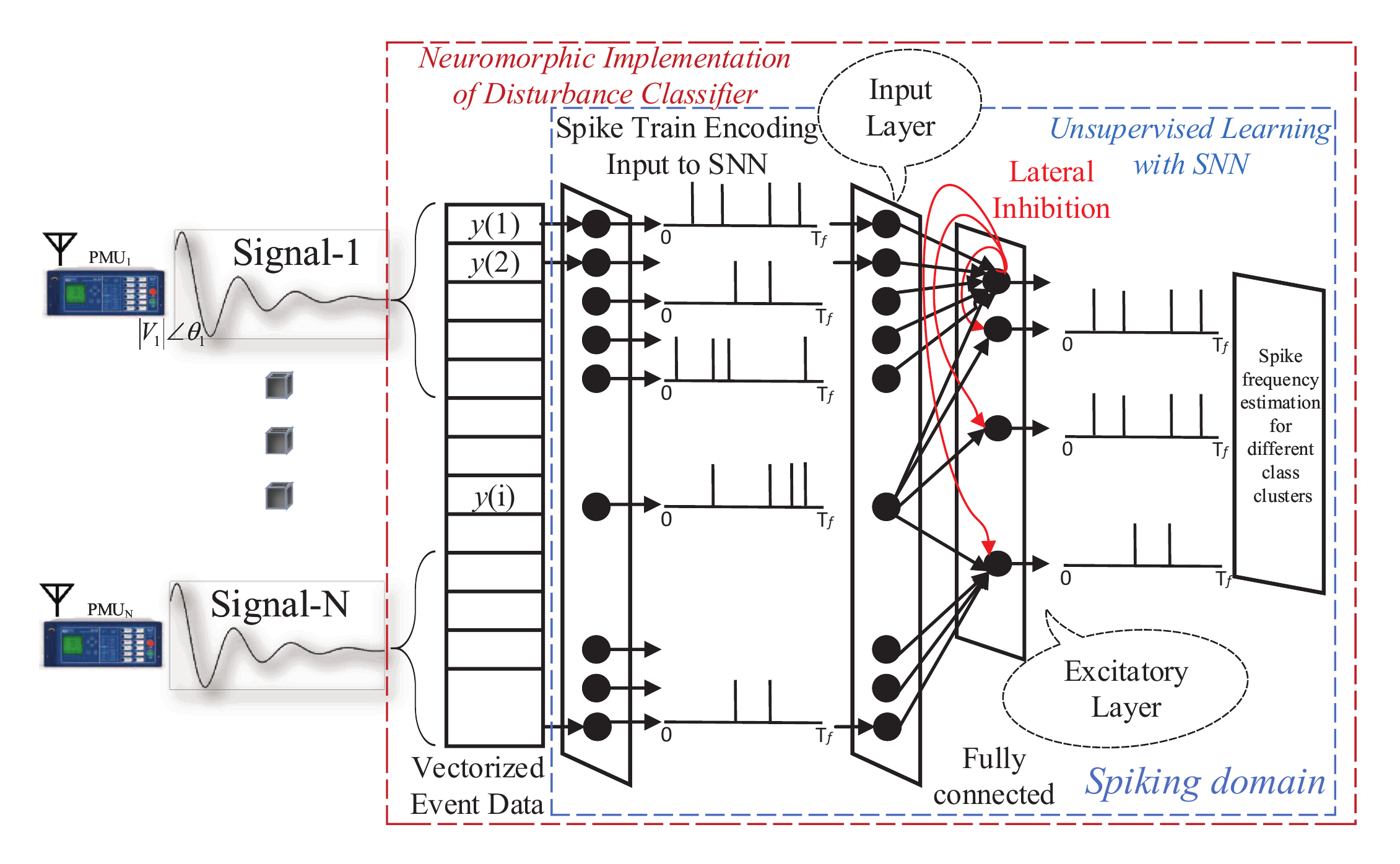}
\vspace{-25pt}
	\caption{Spike train representation of incoming PMU signals communicated to a Spiking Neural Network (SNN) architecture for disturbance event classification.}
	\label{fig:UnSupSNNArch}
\vspace{-15pt}
\end{figure}

\vspace{-4pt}
A typical architecture of spiking neural network\cite{sengupta2016hybrid} consists of input layer and excitatory layers as shown in Fig.~\ref{fig:UnSupSNNArch}. The number of neurons present in input layer is a function of number of signals and corresponding samples of the data window. Each neuron in the input layer is fully connected to the neurons in the excitatory layer. A connection between two neurons is called a synapse. After receiving the spike trains from input layer, excitatory postsynaptic potentials are generated. The synaptic weights between input to excitatory layer are then trained based on time difference between pre- and post-synaptic spike occurrences through a process called spike time dependent plasticity (STDP), which promotes temporal correlation in the firing activities of the connecting neurons. The excitatory layer has inhibitory connections in which information from each excitatory neuron is simultaneously backward-propagated to other neurons in the excitatory layer. This is called ``lateral inhibition" whose purpose is to discourage simultaneous firing of multiple excitatory neurons and promotes competition among neurons for learning different input patterns. The firing potential or threshold of each neuron in the excitatory layer is increased (referred to as `Homeostasis') every time it spikes to prevent single neuron from dominating the firing pattern. Each excitatory neuron learns to represent a class and spikes at a higher frequency as compared to other neurons when examples of that class are presented as input to the SNN.

As shown in Fig.~\ref{fig:UnSupSNNArch}, the input layer accepts samples of measurement data window and produces spike train as output, which in turn is sent to the excitatory layer. The synapse response in terms of excitatory postsynaptic potential (EPSP) for a spike train input to a neuron in the excitatory layer can be described as follows.


\vspace{-15pt}

\vspace{0pt}
\begin{equation}
\vspace{-5pt}
EPSP\left( \texttt{t} \right) = \left\{ {\begin{array}{*{20}c}
   {1,} & {\texttt{t}_c  \le \texttt{t} \le \texttt{t}_d {\kern 1pt}, {\kern 1pt} \texttt{s}(\texttt{t}_c ) = 1}  \\
   0 & {otherwise}  \\
\end{array}} \right.
\end{equation}

\noindent The temporal dynamics of membrane potential $u_{post}$ of a leaky integrate-and-fire (LIF) neuron in the excitatory layer in response to EPSP is given by
\vspace{-5pt}
\begin{equation}
\vspace{-5pt}
\tau_{mt} \frac{{du_{post}(\texttt{t}) }}{{d\texttt{t}}} =  - u_{post}(\texttt{t})  + w(\texttt{t})\sum\limits_{\texttt{T}_f } {{\rm EPSP(\texttt{t})} - I} 
\end{equation}

\noindent where, $\tau _{{\rm epsp}}$ and $\tau_{mt}$ are the EPSP and membrane time constants, respectively. $u_{post}$ is being reset after reaching a threshold $u_{th}$ and results in spiking of post-synaptic neuron in the excitatory layer. Immediately after this event, $u_{post}$ is not allowed to change for a refractory period. In addition, Homeostasis was adopted to prevent single neuron from dominating the response pattern, where $u_{th}$ for the spiking neuron is updated to a higher value at a rate of $\theta _r$ as follows.
\vspace{-5pt}
\begin{equation}
\vspace{-5pt}
u_{th}  \Leftarrow u_{th}  + \theta _r u_{th} 
\end{equation}

In this framework, a biologically plausible unsupervised learning mechanism based on STDP \cite{sengupta2016hybrid} is presented. STDP is a form of Hebbian learning, which supports event-driven learning and weight updating only upon arrival of post synaptic spikes during learning phase and can provide low power on-chip learning. The main idea behind STDP is the adjustment of synaptic weights based on the the temporal correlation between the pre- and post-synaptic spike occurrences. STDP-based weight updation for each connection between input to excitatory layer is described as follows.

\vspace{-5pt}
\begin{equation}
\vspace{-5pt}
\Delta w = \left\{ {\begin{array}{*{20}c}
   {\varsigma _ +  e^{\left( {{{ - \Delta \texttt{t}} \mathord{\left/
 {\vphantom {{ - \Delta \texttt{t}} {\tau _ +  }}} \right.
 \kern-\nulldelimiterspace} {\tau _ +  }}} \right)} } & {\Delta \texttt{t} > 0}  \\
   {\varsigma _ -  e^{\left( {{{ - \Delta \texttt{t}} \mathord{\left/
 {\vphantom {{ - \Delta \texttt{t}} {\tau _ -  }}} \right.
 \kern-\nulldelimiterspace} {\tau _ -  }}} \right)} } & {\Delta \texttt{t} < 0}  \\
\end{array}} \right.
\end{equation}

\noindent where, $\Delta \texttt{t} = \texttt{t}_{post} - \texttt{t}_{pre}$. $\texttt{t}_{post}$ and $\texttt{t}_{pre}$ are the pre- and post-synaptic firing time instances, respectively. $\tau_{+}$, $\tau_{-}$, $\varsigma_{+}$, $\varsigma_{-}$ are weight updating constants.

\vspace{-15pt}
\subsection{Learning using Supervised Spiking Neural Network}\label{sec:SNN_Sup_Descrip}
\vspace{-2pt}

Although unsupervised SNN has the advantage of low power on-chip local learning without the requirement of any labelled data, supervised SNN gives much better performance and are currently more scalable to complex pattern recognition tasks \cite{sengupta2019goingdeeper}. Here, we explore a particular type of supervised SNN training, namely ANN-SNN conversion, for power system disturbance classification. In a supervised learning framework, a particular type of sparsely-firing event-driven spiking deep network \cite{ANNSNN_Diehl,RELUfromDeihl,sengupta2019goingdeeper} can be constructed by training an ANN using standard backpropagation and then converting to SNN. This technique exploits the relationship between number of spikes produced by integrate-and-fire (IF) spiking neuron over an interval in SNN and the output activation of Rectified Linear Unit (ReLU) network (ANN). Typically, the output activation $v_j$ of a ReLU neuron $j$ as a weighted sum of neuronal activations received from previous layer ($\forall y_i$) can be expressed as follows.

\vspace{-8pt}
\begin{equation}
\vspace{-5pt}
v_j  = \max \left( {0,\sum\nolimits_i {w_{ij} y_i } } \right)
\end{equation}

\noindent Similarly, in case of IF neurons (without leak and refractory period), input $y_i$ encoded as spike train $s^{y_i}(\texttt{t})$, with expectation $E\left[ {\texttt{s}^{y_i}  (\texttt{t})} \right] \propto y_i$, which is integrated over $\texttt{T}$ to update membrane potential $u_{mem}$, which upon crossing the threshold $u_{th}$ generates output spikes $s^{v_j}$.

\vspace{-8pt}
\begin{equation}
\vspace{-5pt}
u_{mem} \left( {\texttt{t} + 1} \right) = u_{mem} \left( \texttt{t} \right) + \sum\nolimits_i {w_{ij} \texttt{s}^{y_i } \left( \texttt{t} \right)} 
\end{equation}

In this framework, a ReLU-Based Feed-Forward Neural Network is first trained. To ensure minimum loss in accuracy and faster convergence during the conversion process, a weight normalization is adopted \cite{ANNSNN_Diehl,RELUfromDeihl,sengupta2019goingdeeper} to optimize the ratio of neuron thresholds to synaptic weights. This is achieved by propagating the training set through the network and recording the maximum ReLU activations per layer. The weights are then scaled according to the maximum possible activation within the training set \cite{ANNSNN_Diehl}. The final weight-normalized spiking network can be tested for classification of disturbances in the test set. For an extensive discussion on spiking neuron models and network architectures, readers are directed to reference \cite{taherkhani2020review}.

\begin{figure}
	\centering
	\hbox{\hspace{10pt}\includegraphics[trim={0.4cm 1.1cm 1cm 0.6cm},clip,width=0.48\textwidth]{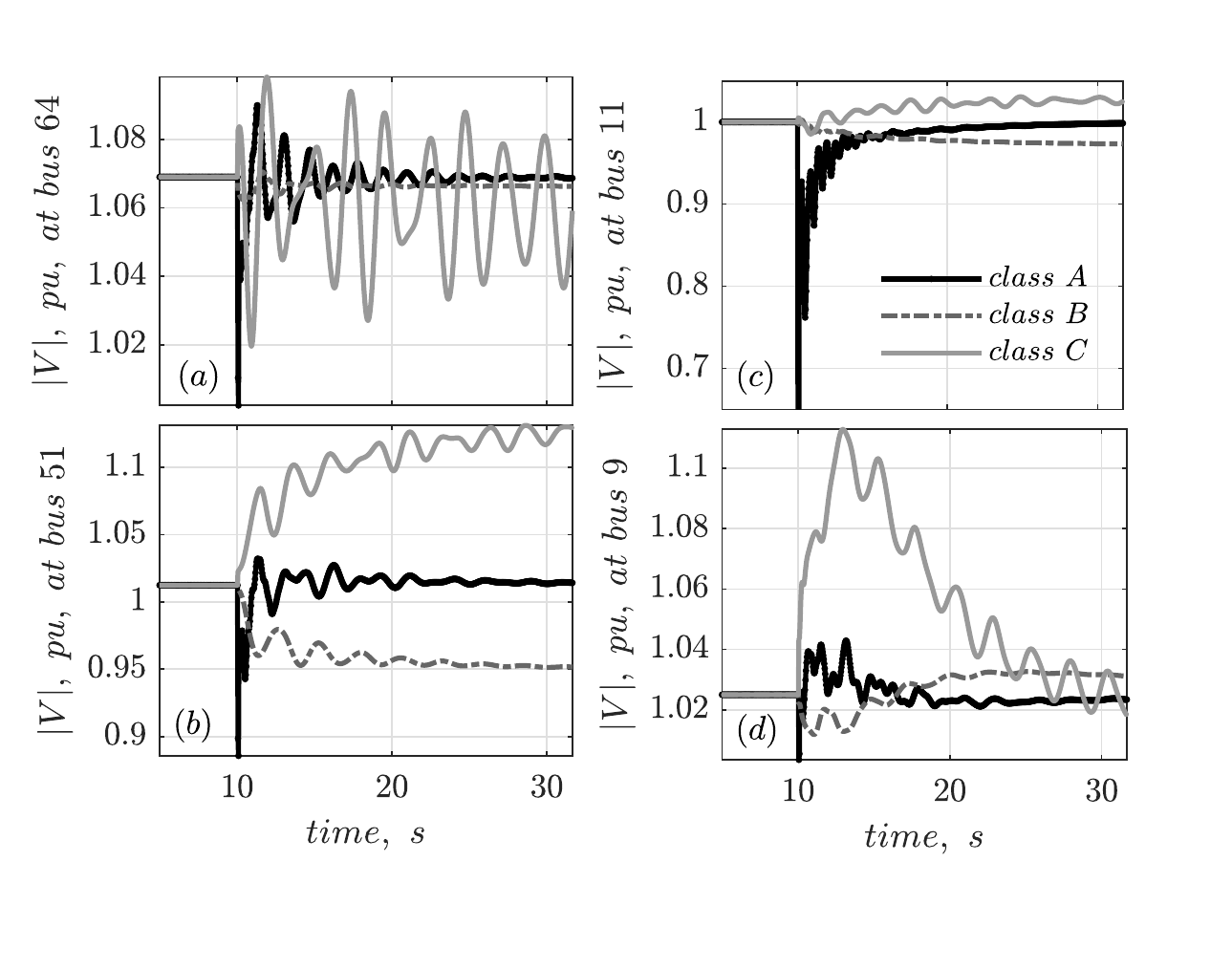}}
\vspace{-10pt}
	\caption{Trajectories of 4 voltage measurements ($|V|, pu$) at buses 64, 51, 11, 9 during disturbance applied at 10s. Three traces show the type of disturbance; Class-A: bus fault, Class-B: generator outage, Class-C: load trip cases.}
\vspace{-11pt}
	\label{fig:TimSerDistExampl}
\end{figure}

\begin{figure}
	\centering
	\vspace{-10pt}
	\hbox{\hspace{-8pt}\includegraphics[trim={0.5cm 0.2cm 1cm 0.3cm},clip,width=0.5\textwidth]{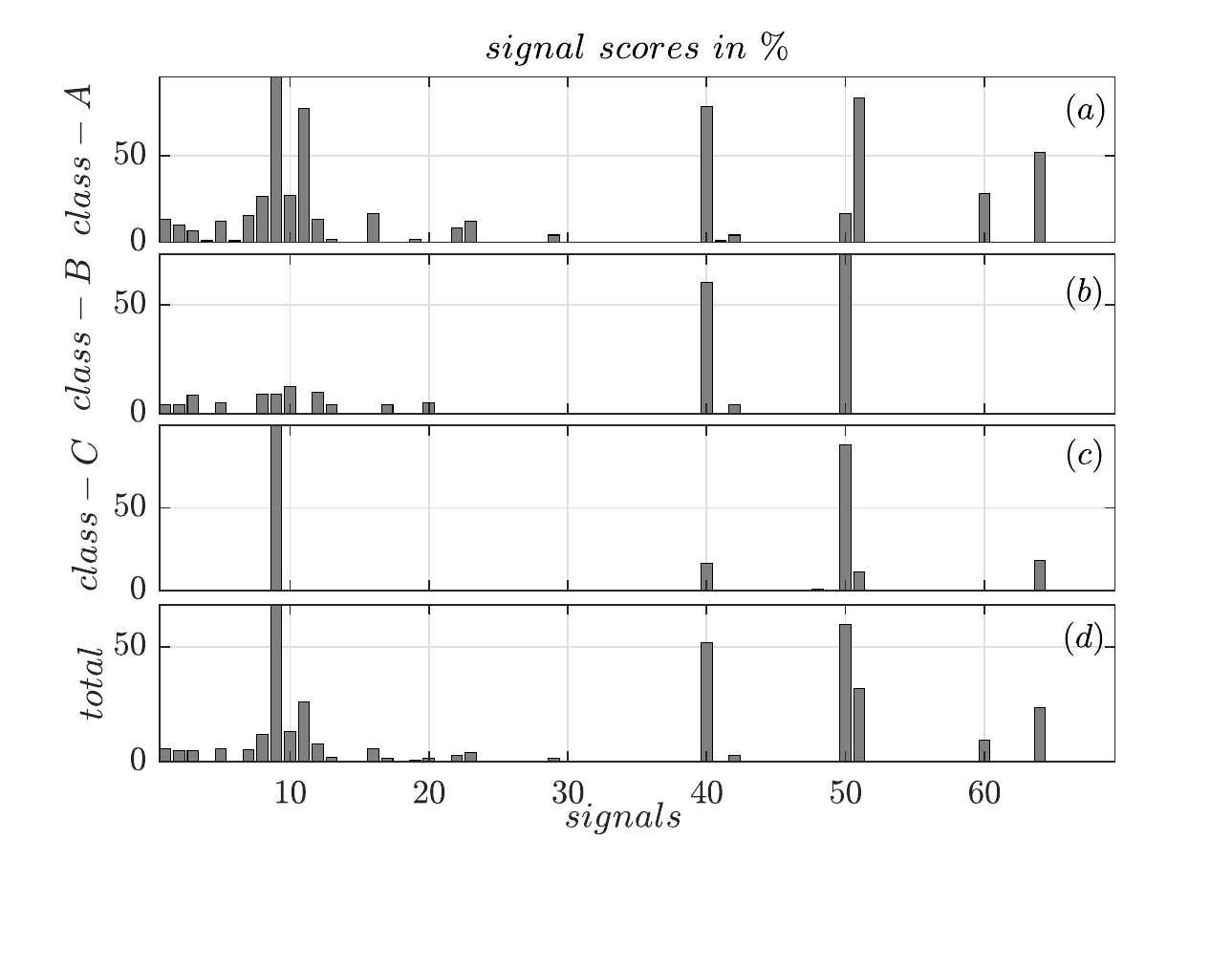}}
\vspace{-32pt}
	\caption{ For 69 voltage magnitude signals $|V|$, signal scores in \% of total number of events was obtained over 3 classes of disturbance events and is presented in (a)-(c). The total score $Sc$ as a sum of the 3 scores was calculated for each signal and is shown in (d). }
\vspace{-15pt}
	\label{fig:SigScoreFig4bar_0}
\end{figure}


\vspace{-14pt}
\section{Results and Performance Evaluation}\label{sec:Results}
\vspace{-4pt}
\subsection{Experimental Setup}\label{sec:X}

A $16$-machine, $5$-area New England-New York system \cite{pal2006robust} 
is considered with PMUs installed at all 69 buses. A disturbance event data set consisting of (1) bus faults (BF), (2) generator outages (GO), and (3) load trip (LT) cases were created via a contingency analysis under $3$ operating conditions with different loading scenarios. During each simulation, disturbance was applied at $10$s and measurements of voltage magnitudes $|V|$ from all buses and frequencies $f$ from $16$ generator terminals were recorded for $40$s at a sampling frequency of $30$Hz. A total of $201$ bus faults, $38$ generator outages, and $156$ load trip cases were generated. Figure \ref{fig:TimSerDistExampl} shows time-domain plots of $4$ voltage magnitude measurement $|V|$ signals recorded under $3$ types of disturbance cases.

\tiny{$\blacksquare$} \normalsize \emph{Pre-processing of voltage magnitudes and frequency measurements:} For classification purposes, the entire data set from the three classes were divided into 10-fold cross validation sets with 60\% of the event examples used for training and 40\% for testing. A window size of 15s was considered for extraction from starting of event at 10s while detrending and downsampling to 10 Hz was applied on each signal.
\vspace{-14pt}
\subsection{Signal Selection on Voltage Magnitude Measurements}\label{sec:X}
The QR decomposition-based signal selection algorithm proposed in Section~\ref{sec:SigSelFramework} was applied on the data window $Y$ with all $|V|$ measurements. Figure \ref{fig:SigScoreFig4bar_0} shows the signal importance scores $Sc$ calculated for 69 signals with 3 classes of event examples and their total scores. As can be seen from Fig.~\ref{fig:SigScoreFig4bar_0}, signals $|V_9|$, $|V_{40}|$, $|V_{50}|$ are participating in forming low rank subspace of more number of events. A group of $\hat n$ signals with the highest total scores were selected from set $S$. An estimation error formula is presented next for performance evaluation of the proposed heuristics in preserving the information in the entire data set. 

Given $m$ signals with $N~(>>2m)$ samples from each training event example $Y_h$, suppose the proposed heuristic selects $\hat n$ signals of higher importance scores present in $S$ (with the rest signals in set $\bar S$). Then, given $pN~(>>m)$ samples of all $m$ signals, where $p~(\ge 0.5)$ shows a fraction of $N$ samples, the rest $(1-p)N$ samples of $(m-\hat n)$ signals in $\bar S$ can be estimated using $(1-p)N$ samples of $\hat n$ signals in $S$ with a least squares (LS) estimation procedure as follows.

\begin{itemize}
    \item Given training event examples $Y_h, h \in [1:H_{train}]$
    \item Initialize sets $S$, $\bar S$, $D$, $\bar D$.
    \item $\textit{while}$ $h \le H_{train}$ (for each $Y_h$)
    \begin{enumerate}
    \item Calculate $\beta  = \left( {Y^{(S,D)}} \right) ^{\dagger} {Y^{(\bar S,D)}}$; $\dagger$ indicates pseudo-inverse of a matrix.
    \item Estimate ${\hat Y^{(\bar S,\bar D)}}=  {Y^{(S,\bar D)}} \beta$
    \item  Calculate LS estimation error for example event $Y_h$;
    $e_h =||{Y^{(\bar S,\bar D)}}-{\hat Y^{(\bar S,\bar D)}}||_2$
    \end{enumerate}
    \item Calculate mean error $\mu _{X}$ across examples $h \in [1,H_{train}]$
\end{itemize}

Here ${Y^{(S,D)}}$ denotes submatrix of $Y_h$ with signals present in set $S$ and samples present in set $D$. Subscript $h$ in notation ${Y^{(,)}}$ is dropped for simplicity. $D$ and $\bar D$ show the samples in the range $[1, pN]$ and $[pN+1, N]$, respectively. Using the above steps, an estimation error $\mu_X$ can be calculated for $H_{train}$ example events. 

\begin{figure}
	\centering
	\vspace{-20pt}
	\hbox{\hspace{-8pt}\includegraphics[trim={0.5cm 0.2cm 1cm 0.4cm},clip,width=0.5\textwidth]{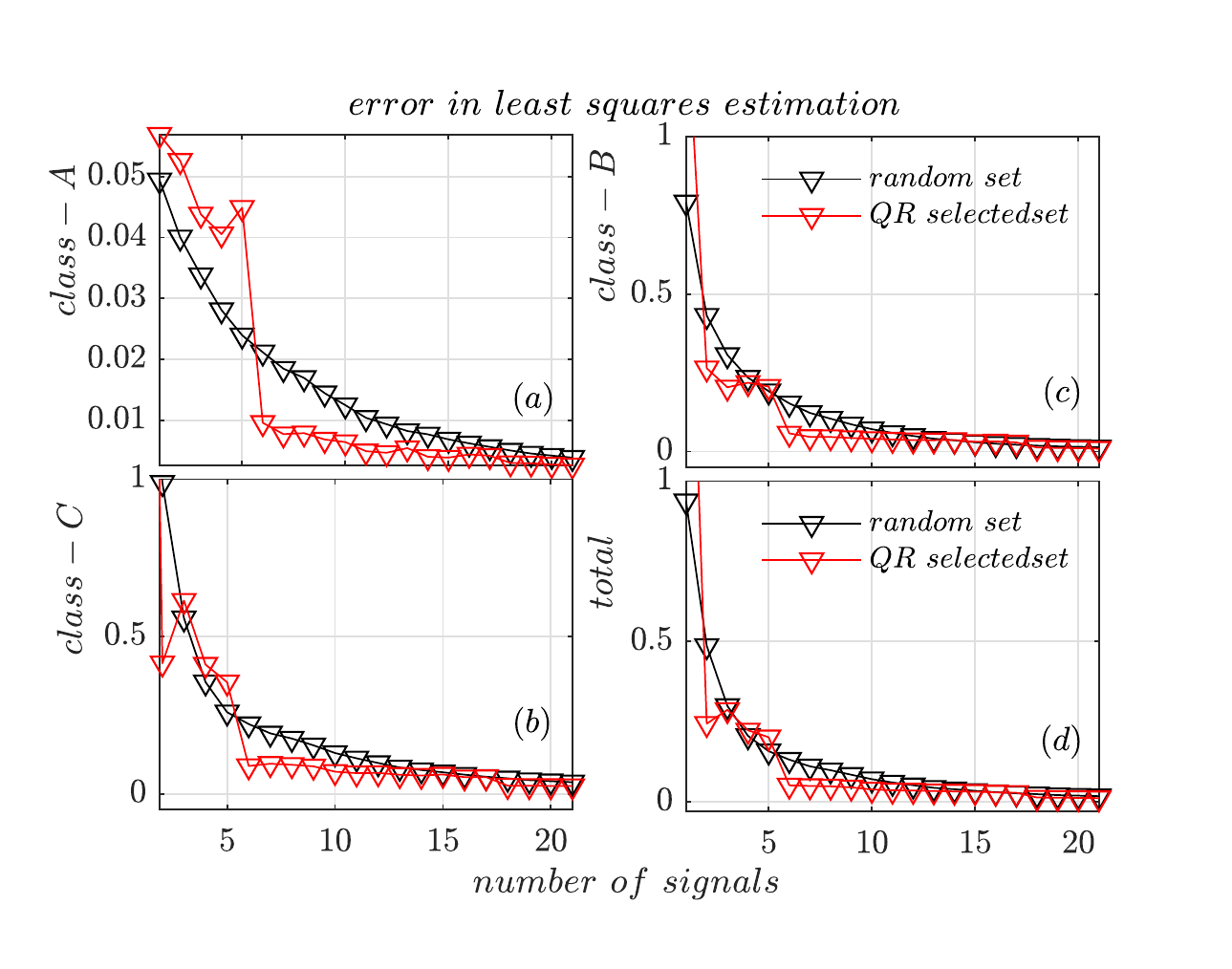}}
\vspace{-21pt}
	\caption{Average error in LS estimation of signals in the set $\bar S_{W_r}(\bar S)$ using the selected signal set $S_{W_r}(S)$, when the number of signals ($|S_{W_r}|(|S|)$) varies from 1-21 as shown in $x$-axis.}%
\vspace{-15pt}
	\label{fig:LSEstim_paper0}
\end{figure}

A Monte-Carlo simulation was then conducted for 100 randomly generated signal combinations $\textbf{W}$, each with $\hat n=21$ number of signals such that for each randomly generated set $W_r \in \textbf{W}$, $|W_r|=\hat n=21$. The effect of considering selection of different number of signals in set $S_{W_r} \subset W_r, (\bar S_{W_r} =W_r-S_{W_r})$, and each with multiple combination of signals on the estimation error for all disturbance events was then compared against the proposed set $S$ obtained using the proposed heuristics. For each combination of signals in each random set $W_r$, the estimation errors were calculated by considering $|S_{W_r}|=1~(|\bar S_{W_r}|= 69-|S_{W_r}|)$, and in each subsequent iteration, $|S_{W_r}|$ was increased by 1 by including signals from the combination in $W_r$. Error $\mu_X$ in estimation of the unselected signals in $\bar S_{W_r}(\bar S)$  is calculated as a function of set $S_{W_r}(S)$ and the number of signals $|S_{W_r}|(|S|)$ for each signal set $W_r$ and the proposed set of signals in $S$ -- see, Fig.~\ref{fig:LSEstim_paper0}. A decrease in the estimation errors is observed by including more number of signals in set $S_{W_r}(S)$. This reduction is more significant for proposed set $S$ as compared to any randomly selected set $S_{W_r}$. This shows the ability of the proposed set $S$ to contain most information present in the entire data set. Therefore, in this work, $|S|=\hat n=10$ number of signals in $S$ were selected to provide input to event detection and classification algorithms.

\begin{figure}
	\centering
	\vspace{-12pt}
	\hbox{\hspace{-8pt}\includegraphics[trim={0.5cm 0.2cm 1cm 0.4cm},clip,width=0.5\textwidth]{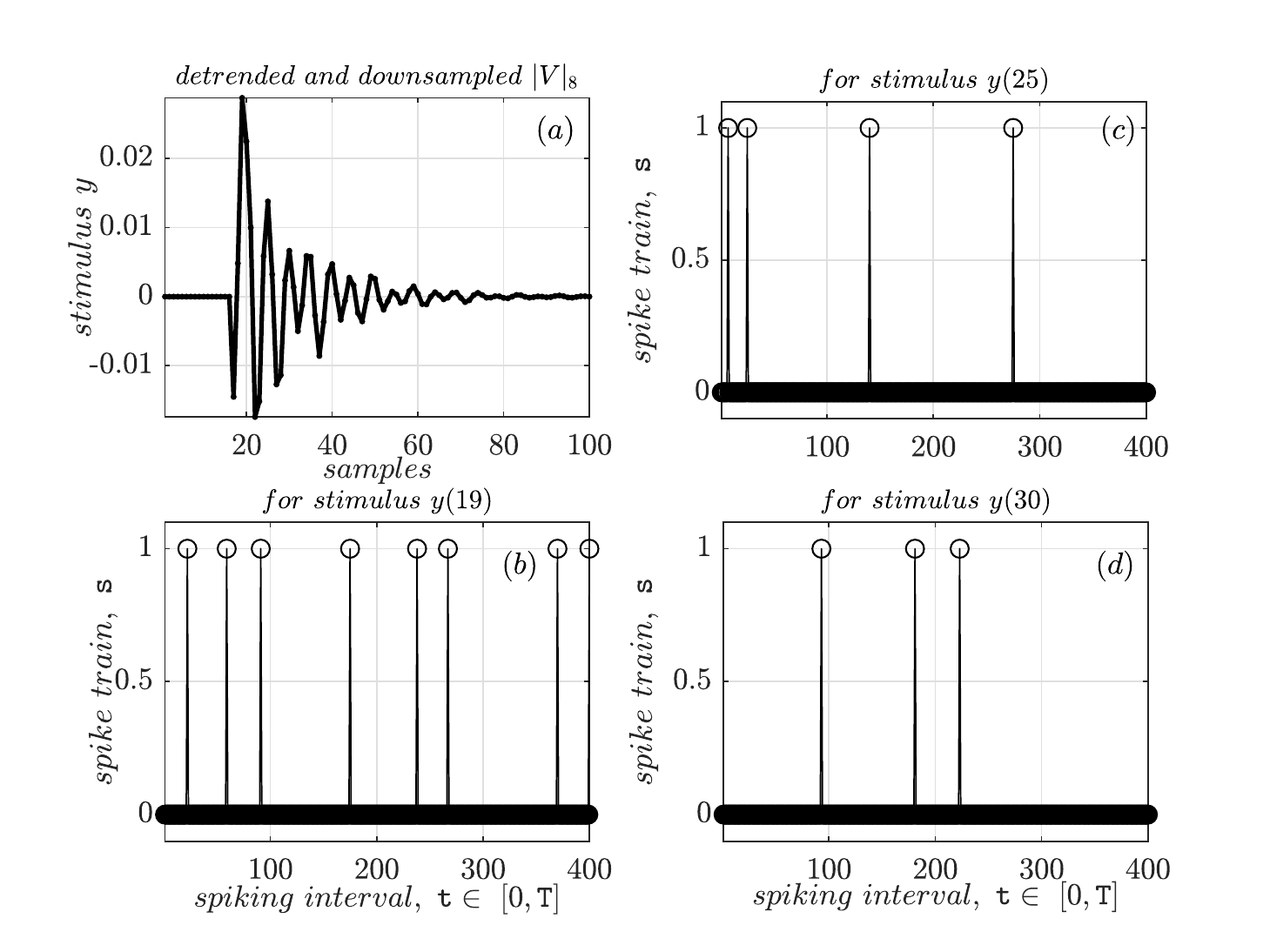}}
\vspace{-10pt}
	\caption{Preprocessed voltage signal $|V_8|$ as stimulus $y$ presented as input to SNN is shown in (a). (b)-(d) shows spike train representation of samples $y(19), y(25), y(30)$, respectively over a spike train duration of 400 samples.}
\vspace{-14pt}
	\label{fig:SpikingRepsnFig0}
\end{figure}

\vspace{-10pt}
\subsection{Spike Train Representation}\label{sec:SpikeTrainRepp}

In this step, the entire window of data matrix for each event example was converted to a 1-D vector by stacking the signal samples present in the window, which is further encoded as Poisson spike trains with an average spiking rate proportional to the maximum value of the data window. Figure \ref{fig:SpikingRepsnFig0}(a) shows the detrended and downsampled version of voltage magnitude measurement signal from bus 8. The corresponding spike train representation for $19^{th}$, $25^{th}$, and $30^{th}$ samples are shown in Fig.~\ref{fig:SpikingRepsnFig0}(b), (c), and (d), respectively. Spike train generated for each example event data was then presented as input to SNN.

\subsection{SNN's Classification Performance Analysis}\label{sec:SNNClassf}

The experiment starts with training and inference on 10 cross-validation (CV) sets in the following supervised and unsupervised learning frameworks -- $1)$ Unsupervised learning framework: $a)$ $U_1$: Self-organising maps (SOM); $b)$ $U_2$: LIF based SNN; $2)$ Supervised learning framework: ANN-SNN  conversion based SNN. The following sections present classification results obtained with supervised and unsupervised learning techniques.

\begin{table*}[b]
\vspace{-20pt}
	\centering
	\footnotesize
\vspace{0pt}
	\renewcommand{\arraystretch}{0.9}
	\vspace{0pt}
	\caption{Comparison of classification accuracy between SOM ($U_1$) and SNN ($U_2$) with unsupervised learning using $|V|$ and $f$ features}
	\vspace{-6pt}
	\label{Table_InII}
	\begin{tabular}{c||c||c|c|c|c|c|c|c|c|c|c|c}
		\hline
		\bfseries  Feature type & CV Sets &  1 &  2 &  3 &  4 &  5 &  6 &  7 &  8 &  9 &  10 & Average accuracy \\
		\hline
		\bfseries Voltage $|V|$ & Training (SOM) &  93.6 &	96 &	93.7 &	96.5 &	97 &	96.6 &	95 &	92.8 &	92.5 &	96 &	94.9  \\
		\hline
		\bfseries Voltage $|V|$ & Test (SOM) &  93.3 &	93 &	96 &	92 &	92 &	91.6 &	89 &	87.5 &	92.1 &	98 &	92.4  \\
		\hline
		\bfseries Voltage $|V|$ & Training (SNN) &  90 &	87 &	90 &	90.3 &	91.5 &	85 &	86 &	90.1 &	86 &	90.1  & 88.6 \\
		\hline
		\bfseries Voltage $|V|$ & Test (SNN) &  92.5 &	92 &	92 &	93 &	94.6 &	91 &	94 &	94.2 &	93 &	93  & 92.9 \\
		\hline
		\bfseries Frequency $f$ & Training (SOM) &  96 &	94 &	93 &	93.4 &	93.1 &	97 &	92 &	96.4 &	95 &	94.2 &	94.4  \\
		\hline
		\bfseries Frequency $f$ & Test (SOM) &  95 &	88 &	96.2 &	95.8 &	98 &	90 &	85 &	95.4 &	96.2 &	95 &	93.4  \\
		\hline
		\bfseries Frequency $f$ & Training (SNN) & 93 & 92.3 & 94 &  95.31 & 92.8 & 94 & 97 & 93.7 & 92.6 & 96.14 & 94.08 \\
		\hline
		\bfseries Frequency $f$ & Test (SNN) &  98 & 96.7 &	99.2 &	98.7 &	95.8 &	95 & 99 & 93.7 & 95.4 &	96.7 & 96.82 \\
		\hline
	\end{tabular}
	\vspace{-12pt}
\end{table*}


\begin{table*}[t]
\vspace{-20pt}
	\centering
	\footnotesize
	\renewcommand{\arraystretch}{0.9}
	\vspace{0pt}
	\caption{Classification accuracy of SNN obtained using ANN-SNN based conversion (S1) for supervised learning}
	\vspace{-6pt}
	\label{Table_IIIN}
	\begin{tabular}{c||c|c|c|c|c|c|c|c|c|c|c|c}
		\hline
		\bfseries Feature type & CV Sets &  1 &  2 &  3 &  4 &  5 &  6 &  7 &  8 &  9 &  10 & Average accuracy \\
		\hline
		\bfseries Voltage $|V|$ & Training & 98 & 99 & 98 & 98.6 & 98 & 99 & 99 & 97.3 & 98.3 & 99 & 98.42\\
		\hline
		\bfseries Voltage $|V|$ & Test &  97 &	97 &	97.1 &	95.8 &	95 &	97.9 &	97.2 &	91.2 &	97.2 &	98 &	96.34    \\
		\hline
		\bfseries Frequency $f$ & Training & 98.3 &	99.5 &	99 &	98 &	97.8 &	99 &	99 &	97.3 &	97.6 &	99.1 &	98.4 \\
		\hline
		\bfseries Frequency $f$ & Test & 98 &	91.2 &	96.2 &	89 &	91 &	96.7 &	90 &	94.6 &	91.2 &	98 &	93.6   \\
		\hline
	\end{tabular}
\end{table*}

\begin{table}[b]
\vspace{-18pt}
	\centering
	\footnotesize
\vspace{0pt}
	\renewcommand{\arraystretch}{0.9}
	\vspace{0pt}
	\caption{Comparison of performance measures between ANN and SNN}
	\vspace{-2pt}
	\label{Table_Ext_1}
	\begin{tabular}{c||c|c}
		\hline
		\bfseries For each MAC (ANN)  &  Unsupervised &  Supervised  \\
		\bfseries to AC (SNN) operation & Learning &  Learning  \\
		\hline
		\bfseries Ratio of number of operations & 1.2  & 1.1	 \\
	    \hline
		\bfseries Ratio of Energy consumption & 6.12  &	5.61 \\
		\hline
		\bfseries \% of Energy saving with SNN & 83.67  &	82.18  \\
		\hline
	\end{tabular}
\end{table}


\subsubsection{Unsupervised Learning}\label{sec:Res_Classf_Unsup}

In this framework, a network of size $[\hat nN,10]$ is presented, where $\hat nN$ represents the number input neurons as well as the number elements in the 1-D vector, which leads to $\hat nN =1090$ for $|V|$ signals and $\hat nN =2512$ for $f$ signals. The number of neurons in output (excitatory) layer of SOM (SNN) is kept 10. The average accuracy for event detection with an SNN on 10 CV tests is 96.5\% for $|V|$ and 97.7\% for $f$ signals, which ensures unsupervised online event-driven operation. The accuracy for disturbance event classification is shown in Table \ref{Table_InII}. The average accuracy on the test set is 92.4\% with $U_1$ and 92.9\%  with $U_2$ for $|V|$ and 93.4\% with $U_1$ and 96.8\% with $U_2$ for $f$ signals.

\subsubsection{Supervised Learning}\label{sec:sec:Res_Classf_Sup}
In this framework, a ReLU based ANN-SNN of size $[\hat nN, h_1, h_2, r]$ as described in Section~\ref{sec:SNN_Sup_Descrip}, is considered for supervised classification using voltage and frequency information of the simulated events with $\hat nN$ input nodes, two hidden layers each with $h_1,h_2$ nodes, respectively and $r$ output nodes with all fully connected layers. Maximum number of epochs was kept to 30 and a mini batch size of 100 was adopted during training ANN through standard backpropagation. The weights from the ReLU network is then scaled to obtain the SNN version of the trained ANN. The results obtained with ANN-SNN conversion are presented in Table \ref{Table_IIIN}, which shows the average classification accuracy is 96.3\% for voltage magnitudes and 93.6\% for frequency measurements and proves the effectiveness of SNN in supervised classification tasks. The following sections present a discussion on the computational performance analysis of both ANN and SNN algorithms.

\vspace{-16pt}
\subsection{Evaluation of Computational Burden}
\subsubsection{Unsupervised Learning}\label{sec:Res_Comp_Unsup}

Although an estimation of the entire hardware overhead or actual energy consumption is outside the scope of the work, a comparison of number of computations per synaptic operations with techniques $U_1$ and $U_2$ are presented based on the number of multiply-accumulate (MAC) operation. A typical ANN with $i$ input neurons and $r$ output neurons requires $ir$ multiplication and addition operations with one forward pass of the fully connected feed forward network. In contrast, SNN performs only accumulate (AC) operations over the entire spiking interval $[0,\texttt{T}]$ in layer $j$ only upon receiving a spike from previous layer $j-1$. Number of AC operation per layer $j$ becomes a product of average number of accumulated spikes per neuron $s_{m}(j)$ from previous layer $j-1$ during the spiking interval $\texttt{T}$ and the number of synaptic operations ($ir$) between two consecutive layers. The ratio of number of MAC operations in SOM to AC operations in SNN is estimated to be $\approx 1.2$, which shows a 20\% reduction in number of computations with SNN.

\subsubsection{Supervised Learning}\label{sec:Res_Comp_Sup}

After ANN-SNN conversion, the total number of computations in SNN during testing becomes $\hat nNh_1s_{m}(2)+h_1h_2s_{m}(3)+h_2rs_{m}(4)$, which is much less as compared to $\hat nNh_1+h_1h_2+h_2r$ operations involved with the trained ANN in the process. The ratio the number of MAC operations involved in ANN to the AC operations in SNN is $\approx 1.1$. This ratio is expected to further increase with more layers in networks~\cite{sengupta2019goingdeeper}, which is due to exponential increase in neuron spiking sparsity with network depth. This could be very useful at control centers for larger datasets. This proves the capability of spiking neurons in achieving similar  accuracy as conventional techniques in disturbance classification with less computational burden.

Moreover, SNN is more efficient even for the worst case scenarios when an input pattern demands more binary AC operations than real-valued MAC operations, since SNNs can be implemented in energy efficient neuromorphic hardware \cite{PfeifferPfeil,Esser11441}. For instance, an estimated ratio of order of magnitude of energy consumption per MAC operation (with one 32-bit ADD and 32-bit MULT) to AC operation (with 32-bit ADD) is $5.1$ for floating point computations \cite{EnergyRelative}. A summary of projected energy savings for disturbance classification has been presented in Table~\ref{Table_Ext_1}, which shows SNN deployed for disturbance classification operations can save upto 83.6\% of energy which is wasted by ANN in the unsupervised and upto 82\% in supervised framework. This implies by deploying SNNs, which are inherently low energy networks and incurs less computations, the energy savings will significantly increase with deep networks. While an actual estimate is outside the scope of the work, the evidence presented through AC operations encourages implementation of the event-driven classifier in a neuromorphic computing hardware for realization of a spiking environment enabling low energy computations and efficient data processing without sacrificing accuracy. 

The following section demonstrates the potential benefits of deep SNNs over ANNs in a new experiment to show the impact of training with more data at scale and corresponding effects on classification accuracy and computational performance.

\section{Scaling to Deeper Networks}\label{sec:DeepSNN}

 \begin{figure}
 	\centering
 	\vspace{-18pt}
 	\hbox{\hspace{-4pt}\includegraphics[trim={0.2cm 0.04cm 0.8cm 0.6cm},clip,width=0.52\textwidth]{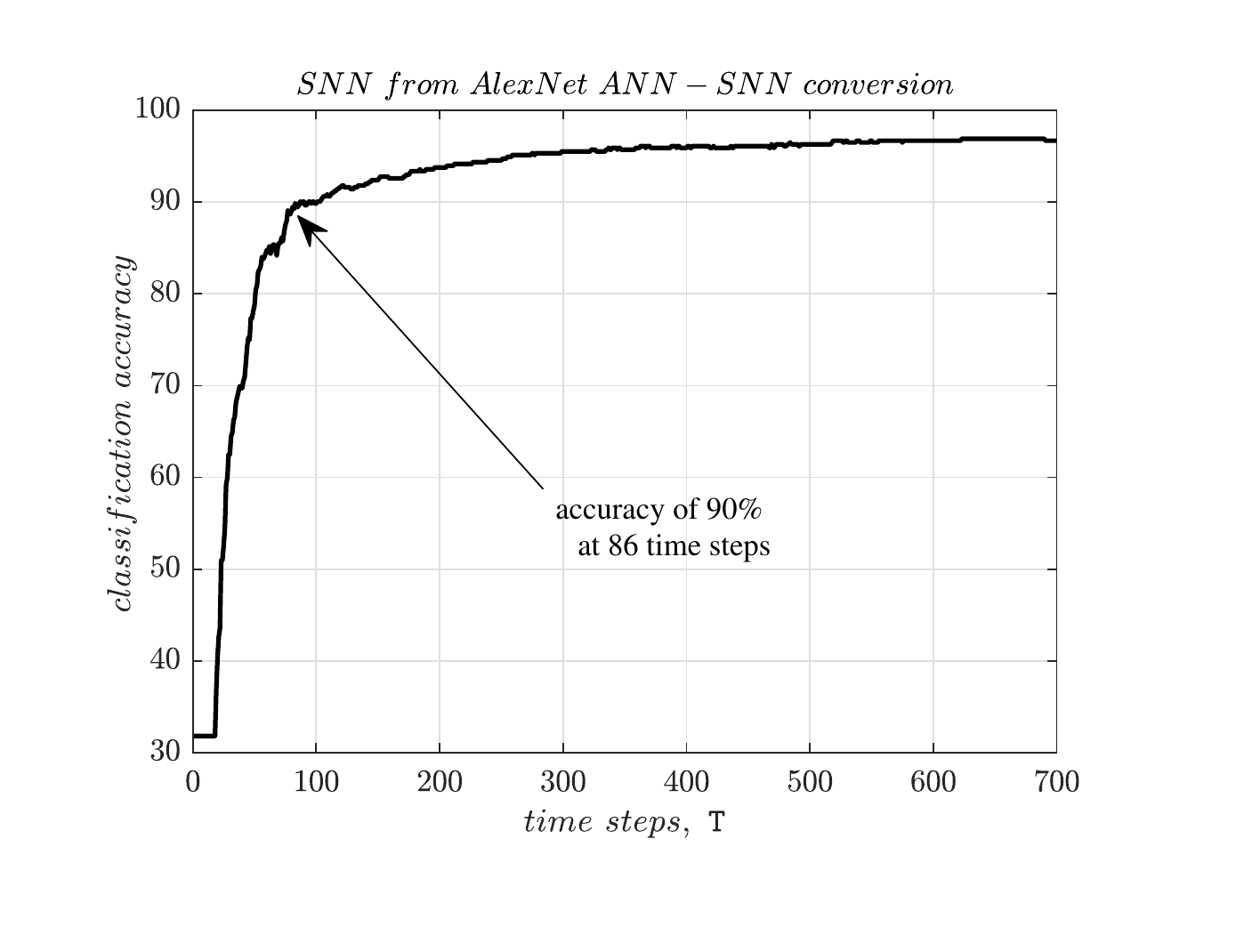}}
 \vspace{-30pt}
 	\caption{The classification accuracy of SNN after AlexNet based ANN-SNN conversion is plotted as a function of time steps~$\texttt{T}$. SNN achieved 90\% validation accuracy at 86 timesteps. The temporal behavior can be exploited for accuracy-efficiency tradeoff.}%
 	\label{fig:dsnn_accuracy}
 \end{figure}

 \begin{figure}
 	\centering
 	\vspace{-10pt}
 	\hbox{\hspace{-2pt}\includegraphics[trim={0.4cm 0.08cm 1cm 0.4cm},clip,width=0.47\textwidth]{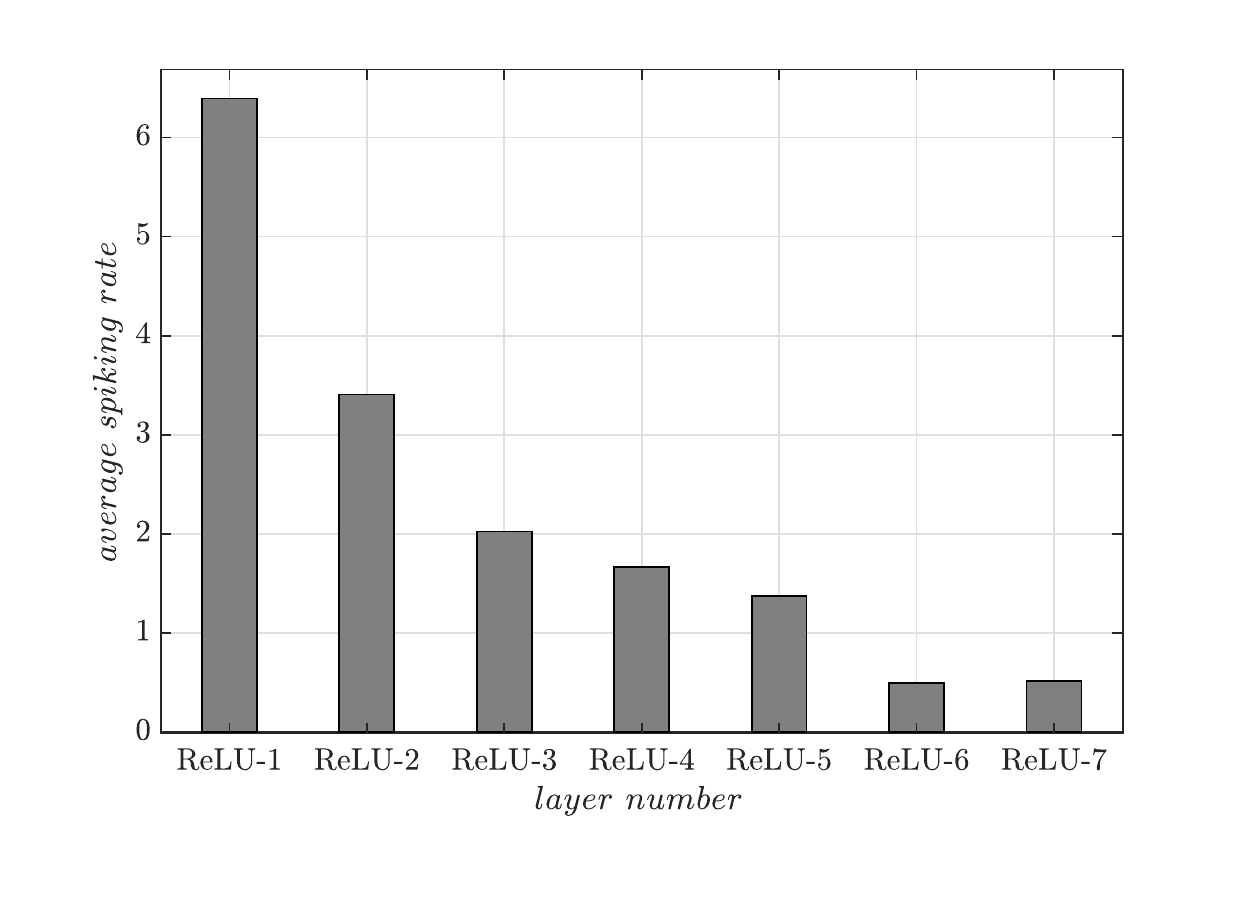}}
 \vspace{-23pt}
 	\caption{Average spiking activity of the converted AlexNet is plotted as a function of network depth. The SNN spiking sparsity significantly increases as the network depth increases. }%
 \vspace{-20pt}
 	\label{fig:dsnn_bars}
 \end{figure}

\subsection{Experimental Setup}

While the results in the previous section have provided motivation for SNN usage in power system disturbance classification, computational burden reduction becomes a significant issue in complex problem spaces where deep networks exceeding a million parameters become necessary to achieve state of the art accuracy. To substantiate the benefits of SNN usage in such scenarios, a new experiment was conducted with additional class data. Due to limitations in the availability of real-time PMU data and an annotated dataset containing information on the type of disturbance in the general power system community, a set of simulations on the existing test system were performed for collecting time series measurements involving transient data from disturbance events. Four classes of events considered in this experiment are: (1) bus faults (BF), (2) generator outages (GO), (3) load trip (LT) and (4) line outage (LO) cases under 7 different operating conditions with varying loading levels across the network. This resulted in 483 BF, 112 GO, 490 LT, 490 LO labeled cases. A 30 seconds of simulation data was extracted from the selected $|V|$ signals and preprocessed for providing inputs to the neural network.

\subsection{Supervised Training Methodology}
Next, we trained a convolutional network architecture by \textit{supervised learning} and subsequently converted that to a spiking network using ANN-SNN conversion. The SNN simulation is performed using a modified version of BindsNet \cite{bindsnet2018}, a PyTorch based package. The AlexNet \cite{NIPS2012_4824} architecture, an 8-layer deep network consisting of 5 convolutional layers and 3  fully connected layers was selected. Without loss of generality, the network was trained on voltage magnitude features. The training can be extended for other features as well. It is worth noting here that the convolutional layers used in our work are 1D, ideal for time-series data processing. The dataset was then randomly split into two parts: 2/3 for training and 1/3 for validation. The network was trained with a learning rate of 0.001 and a batch size of 32, and subsequently converted to spiking mode for inference. To reduce the high inference latency observed in deep SNNs, a fixed percentile (96.4\% in our case) from the maximum activation histogram was used during the weight normalization process. Such a relaxed normalization scheme significantly reduces the timesteps involved in inference without any observable degradation in classification accuracy~\cite{lu2020exploring}. For an extensive discussion on constraints involved in ANN-SNN conversion and design-time/run-time optimizations, readers are referred to reference~\cite{lu2020exploring}. 

\subsection{Advantages with Deeper SNN}

Figure \ref{fig:dsnn_accuracy} depicts the SNN classification accuracy as a function of timesteps~$\texttt{T}$. This temporal behavior, where the accuracy of the network increases over timesteps, can be leveraged for accuracy-efficiency trade-off - appealing for resource-constrained applications. For instance, the SNN achieves 90\% accuracy at 86 time steps and subsequently saturates in terms of accuracy improvement. Hence, computations involved in the SNN can be significantly reduced by small relaxation in the final accuracy requirement. Using the methodology described in the previous section, the SNN is estimated to be $2.71\times$ energy efficient than the corresponding ANN to achieve a final accuracy of 90\%. It is worth mentioning here that SNNs are power efficient due to sparse event-triggered neural operation. However, the power advantage is achieved at the expense of inference delay. 

While instantaneous power consumption is a key metric driving computational cost overhead for resource constrained devices, resultant energy efficiency (delay x power) will be a trade-off between inference latency (impacting delay) and sparsity of the spike train (impacting power). In this particular example, the power efficiency of the SNN is estimated to be $233\times$ in contrast to the ANN since the inference takes 86 timesteps to reach 90\% accuracy. Note that this is a first order estimate as mentioned before based on computational energy requirements. 

Additionally, Fig. \ref{fig:dsnn_bars} demonstrates that the SNN spiking sparsity significantly increases as the network depth increases. This downward trend of spiking rate as layers go deeper implies SNN's even higher power/energy efficiency when implementing deeper architectures such as VGG \cite{VGG16} and ResNet \cite{He_2016}, which are essential to solve more complex problems.


\vspace{-8pt}
\section{Conclusion} \label{sec:Concl}
This work successfully demonstrated the ability of spiking neural networks (SNN) to process PMU signals for online event-driven disturbance classification. SNNs can exploit the sparsity patterns of events to achieve classification accuracy comparable to ANNs while significantly reducing the energy consumption. In addition, a signal selection technique to determine the signals participating in low rank subspace in multiple disturbance event scenarios was proposed, whose effectiveness was demonstrated by reconstructing samples of multiple signals using the selected signals. Performance comparison of ANN and SNN from classification tests on $16$-machine, $5$-area New England-New York system shows that SNN has the capability to achieve 96.8\% disturbance classification accuracy with upto 20\% reduction in number of computations in terms of MAC to AC operations. Event-driven SNNs in unsupervised (supervised) learning framework results in a projected estimate of energy savings upto 83.6\% (82.18\%) with respect to standard ANNs. In addition, our results on deep SNNs also highlighted the fact that dynamic sparsity benefits would be best realized for deeper networks which, in turn, is a necessity for complex machine learning tasks. Hence, the energy and power benefits of SNNs are expected to further improve with increasing complexity of the power grid disturbance classification problem and in real time operational scenarios. This work opens up the new possibility of utilization of neuromorphic computing algorithms and hardware at control centers for real time disturbance monitoring.

\vspace{-10pt}
\section{Appendix} \label{sec:Appendix}
\subsection{Measurement data}
 $y(t)$ can be expressed in terms of the linearized system matrices ($A_{o}$,$B_{o}$,$C_{o}$,$D_{o}$) of the network~\cite{Kundur} at an operating condition ($o$).
 \vspace{-5pt}
 \begin{equation}\label{eq:TimeS}
 \vspace{-5pt}
 \begin{array}{l}
 \bar y(t) = C_{o} \Phi_{o} L_{o} \bar z(t)
 \end{array}
 \end{equation}
 \noindent where, $\Phi_{o}$ is the modal matrix of $A_o$, columns of which contains the right eigenvectors $ \bar \phi _i$ of $A_o$. $\bar z(t)$ represents the complex exponential associated with $n$ eigenvalues $[{\lambda _1},...{\lambda _n} ]$ of the system at the operating condition and can be expressed as
 \vspace{-5pt}
 \begin{equation}\label{eq:expmode}
 \vspace{-5pt}
 \bar z(t) = \left[ {\begin{array}{*{20}c}
   {e^{\lambda _1 t} } & {...} & {e^{\lambda _n t} } 
 \end{array}} \right]^T 
 \end{equation}
 \noindent Each row in matrix $L_o$ contains vector $\bar l= [l_1 l_2 ...l_n]$, in which each element $l_i$ is a function of initial condition $\nabla  x(0)$ and left eigenvectors $\bar \psi _i$ of $A_o$.
 \vspace{-8pt}
 \begin{equation}\label{eq:Lo}
 \vspace{-5pt}
 l_i  = \bar \psi _i \nabla  x(0)
 \end{equation}

 \noindent where, $\Psi_{o}$ contains the left eigenvectors $\bar \psi _i$. Discretized system response $Y$ ($N \times m$) containing $N$ samples for $t = [0, (N-1)Ts ]$ of $m$ measured signals at an operating condition ($o$), can be expressed as follows.
 \vspace{-5pt}
 \begin{equation}\label{eq:yvff}
 \vspace{-5pt}
 Y^T = C_o \Phi _o L_o V = F_o V \Rightarrow Y=V^*{F_o}^*
 \end{equation}

 \noindent where, $V$ is a Vandermonde matrix ($n \times N$) with $n$ distinct complex modes $[v_1, ..., v_n]$ in the $z$-plane inside the unit circle corresponding to eigenvalues $\lambda_i =F_s ln(v_i)$ in continuous domain. Therefore, $V^*$ contains $n$ columns with $n$ complex exponential time series signals $\bar v_i$ representing the trajectories of modal signals.
 \vspace{-10pt}
 \begin{equation}\label{eq:Vandermonde}
 \vspace{-6pt}
 V^*  = \left[ {\begin{array}{*{20}c}
   {v_1^0 } &  \ldots  & {v_n^0 }  \\
     \vdots  &  \ddots  &  \vdots   \\
   {v_1^{N - 1} } &  \ldots  & {v_n^{N - 1} }  \\
 \end{array}} \right]
 \end{equation}

 \noindent The data window $Y$ can be written as a function of $V$ in \eqref{eq:yvff}, where, rows $\bar y(t) \in \mathbb{R}^m$ of $Y$ are a complex transformation of each row $\bar v(t) \in \mathbb{C}^n$ of $V^*$ by transformation matrix ${F_o}^*$ for each $t \in [0, (N-1)Ts ]$. So we can write $\bar y(t)=\chi_{f_o}(\bar v(t))$, where $y(t)$ is the image of $\bar v(t)$ under mapping $\chi_{f_o}$. For distinct $v_i$'s, the columns of $V^*$ forms a basis of a set of linearly independent vectors of complex exponential time series signals and defines the dimension of $Y$ as $n$ when $m>=n$. 
\vspace{-5pt}
 \subsection{Comparison of classification accuracy of SNN with other machine learning techniques}

Due to the ``architectural similarity" between ANN and SNN and the applicability of ANNs in learning from large-scale datasets, ANN is presented for comparison of accuracy and computational efficiency. Nevertheless, the results with other classification methods are presented here.  

\begin{table}[h]
\vspace{-15pt}
	\centering
	\footnotesize
\vspace{0pt}
	\renewcommand{\arraystretch}{0.9}
	\vspace{0pt}
	\caption{Comparison of classification accuracy}
	\vspace{-2pt}
	\label{Table_other_ML}
	\begin{tabular}{c||c|c|c|c}
		\hline
		\bfseries Accuracy (\%)  &  DT &  SVM & RF & SNN  \\
		\hline
		\bfseries  Training ($|V|$)	 & 98.6  & 95.5  & 95.8  & 98.4 \\
	    \hline
		\bfseries  Testing ($|V|$) & 96.6  &  97.01  & 93  & 96.3 \\
		\hline
		\bfseries  Training ($f$) & 98.7  & 99.2  & 94.4  & 98.4  \\
	    \hline
		\bfseries  Testing ($f$) &  96.2 & 95.08  & 94.5  &  93.6  \\
		\hline
	\end{tabular}
	\vspace{-10pt}
\end{table}

Note that the objective here is to attain a similar accuracy as other machine learning methods with reduced computational overhead. For real-time operations, ANNs or deep ANNs are preferable for large data set as compared to other classification techniques due to the capability of ANNs in representing highly nonlinear functions/decision boundaries. The results in Table \ref{Table_other_ML} indicate that SNN has the capability to achieve similar accuracy as other machine learning techniques such as decision tree (DT), support vector machines (SVM), random forest (RF).  

\textit{Remarks:} Note that for large-scale problems, deep neural networks are known to outperform the traditional ML techniques such as SVM, DT, RF etc. in regression or classification tasks. It is a known fact that  traditional ML techniques are not suitable in achieving generalization while learning from large datasets. Reference \cite{ng2017machine} shows how the scale of the problem drives machine learning progress and highlights the limitation of traditional ML techniques. 

While latest deep neural networks present the best platform to learn from large datasets, a significant improvement in computational efficiency can be achieved through the use of alternative neuromorphic SNN architectures. Thus, deep SNNs are proposed for large power system event detection problems, which can achieve similar accuracy as ANNs, while saving significant energy consumption due to its sparse, event-driven, temporal operation.

\vspace{-11pt}
\bibliographystyle{IEEEtran}
\bibliography{J5_for_ArXiv}

\begin{IEEEbiography}[{\includegraphics[width=1in,height=1.1in,clip]{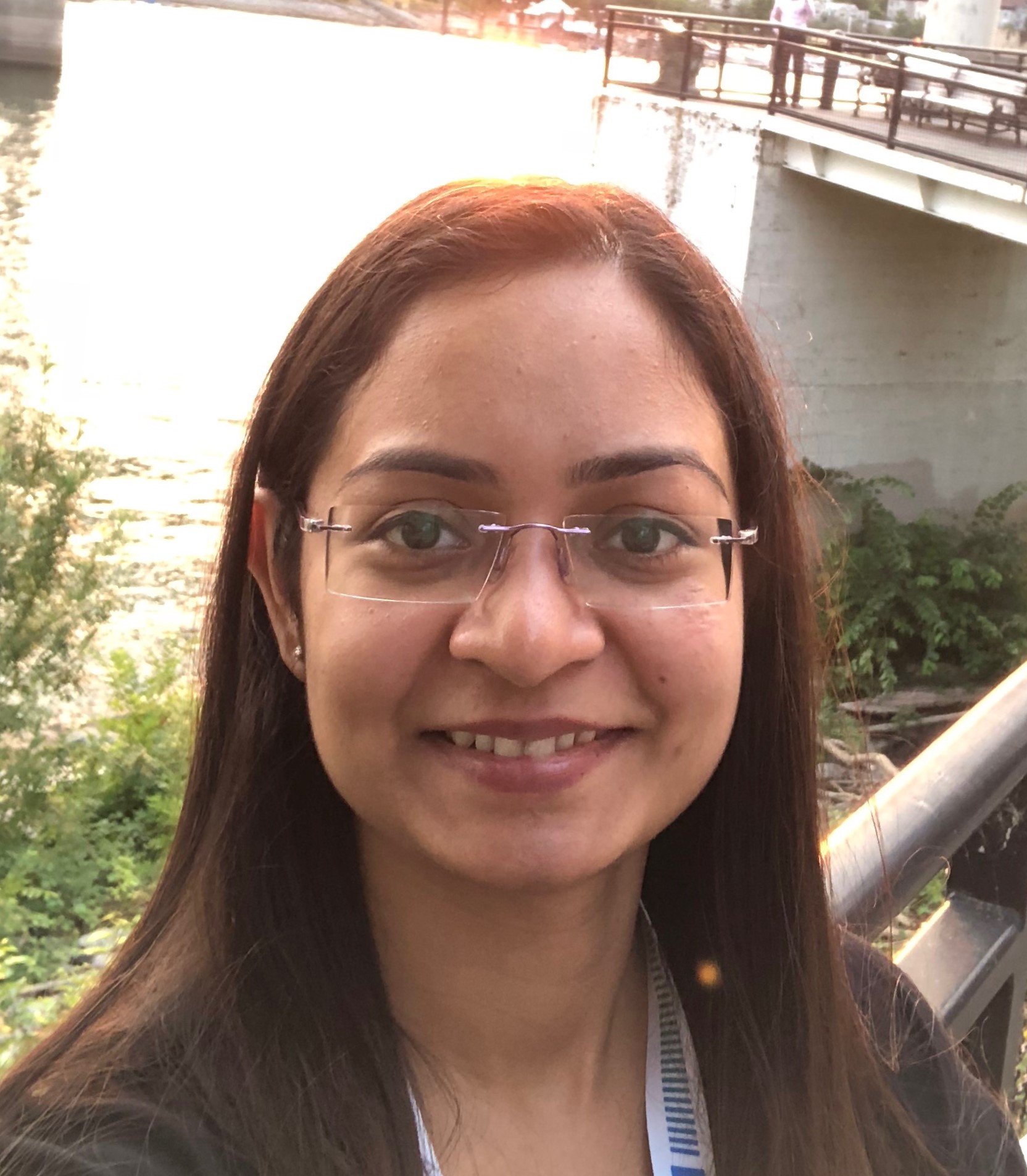}}]{Kaveri Mahapatra} (S’11-S’13-S’20) received her Ph.D. degree from Pennsylvania State University, USA in 2020 in Electrical Engineering. In 2019, she worked with General Electric (GE) Global Research Center, NY, USA as a research fellow intern. She has received her B. Tech. and M. Tech. degree in Electrical Engineering from KIIT University in 2011 and SOA University, India, in 2013, respectively. Presently, she is working as a power systems research engineer with Pacific Northwest National Laboratory, WA, USA. Her current research interests include wide area monitoring, protection and control, power system dynamics and resilience, cyber physical security, machine learning, data analytics and optimization.
\end{IEEEbiography}

\begin{IEEEbiography}[{\includegraphics[width=1in,height=1.1in,clip]{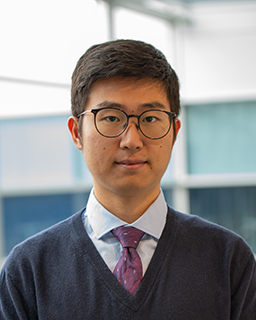}}]{Sen Lu} received the Computer Science Bachelor’s degree from the Department of Computer Science and Engineering, College of Engineering, Pennsylvania State University, University Park, USA, in 2019. 

Sen is currently pursuing a Ph.D. degree in Computer Science and Engineering, in the College of Engineering, Pennsylvania State University, USA. He joined the Neuromorphic Computing Lab of Penn State in 2019. His current research focuses on Computer Vision, Deep Learning, and Neuromorphic Algorithms.
\end{IEEEbiography}

\begin{IEEEbiography}[{\includegraphics[width=1in,height=1.3in,clip,keepaspectratio]{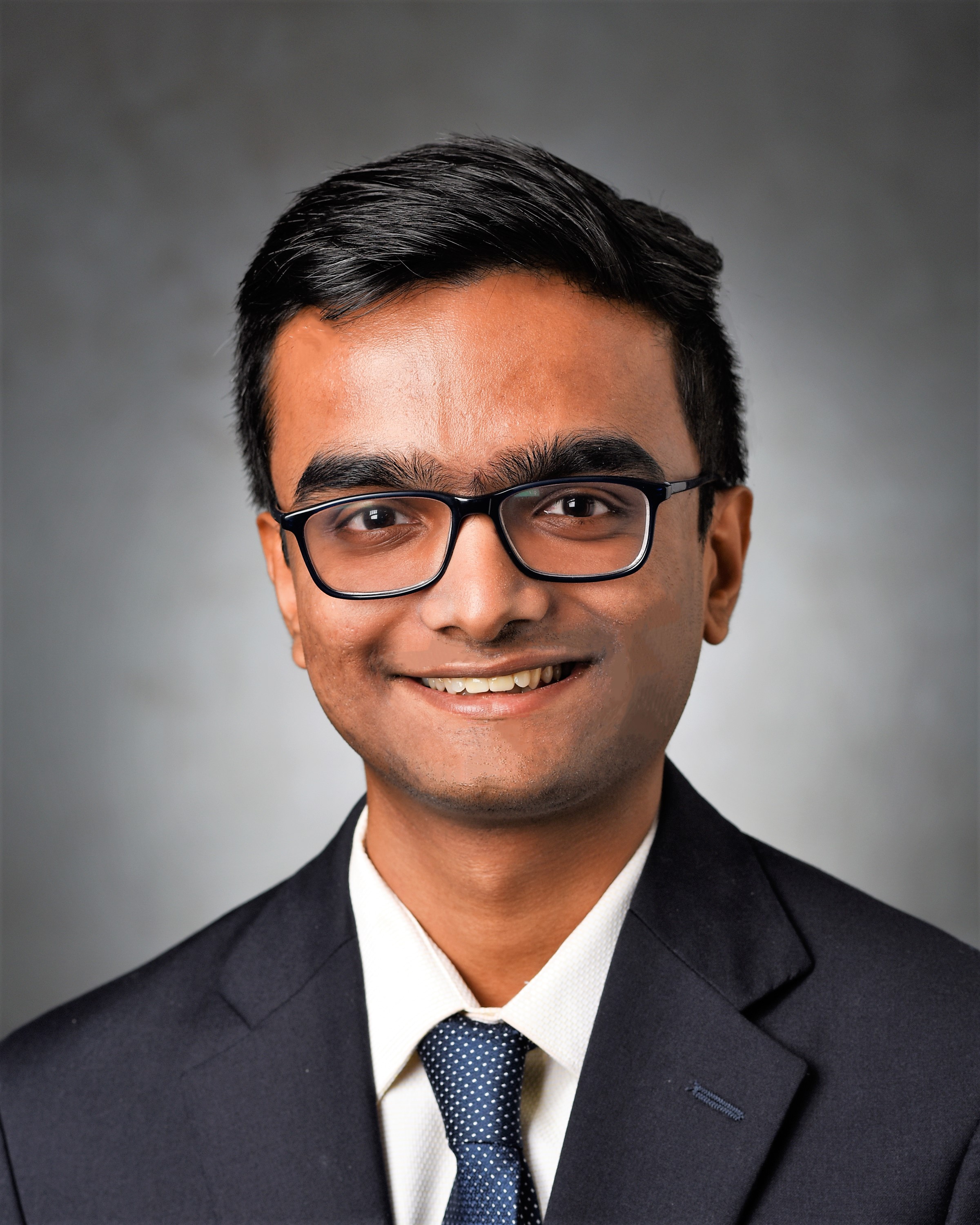}}]
{Abhronil Sengupta}(S’15-M’18)  received the B.E. degree from Jadavpur University, Kolkata, India, in 2013, and the Ph.D. degree in Electrical and Computer Engineering from Purdue University, West Lafayette, IN, USA, in 2018. 

He is an Assistant Professor with the School of Electrical Engineering and Computer Science, Penn State University, University Park, PA, USA. He was as a DAAD (German Academic Exchange Service) Fellow with the University of Hamburg, Hamburg, Germany, in 2012, and as a Graduate Research Intern with Circuit Research Labs, Intel Labs, Hillsboro, OR, USA, in 2016 and Facebook Reality Labs, Redmond, WA, USA, in 2017. He is pursuing an interdisciplinary research agenda at the intersection of hardware and software across the stack of sensors, devices, circuits, systems, and algorithms for enabling low-power event-driven cognitive intelligence. He has published over 70 articles in referred journals and conferences and holds 4 granted/pending U.S. patents. 

Prof. Sengupta was a recipient of the IEEE Circuits and Systems Society Outstanding Young Author Award in 2019, the IEEE SiPS Best Paper Award in 2018, the Schmidt Science Fellow Award Nominee in 2017, the Bilsland Dissertation Fellowship in 2017, the CSPIN Student Presenter Award in 2015, the Birck Fellowship in 2013, and the DAAD WISE Fellowship in 2012.

\end{IEEEbiography}

\begin{IEEEbiography}[{\includegraphics[width=1in,height=1.3in,clip,keepaspectratio]{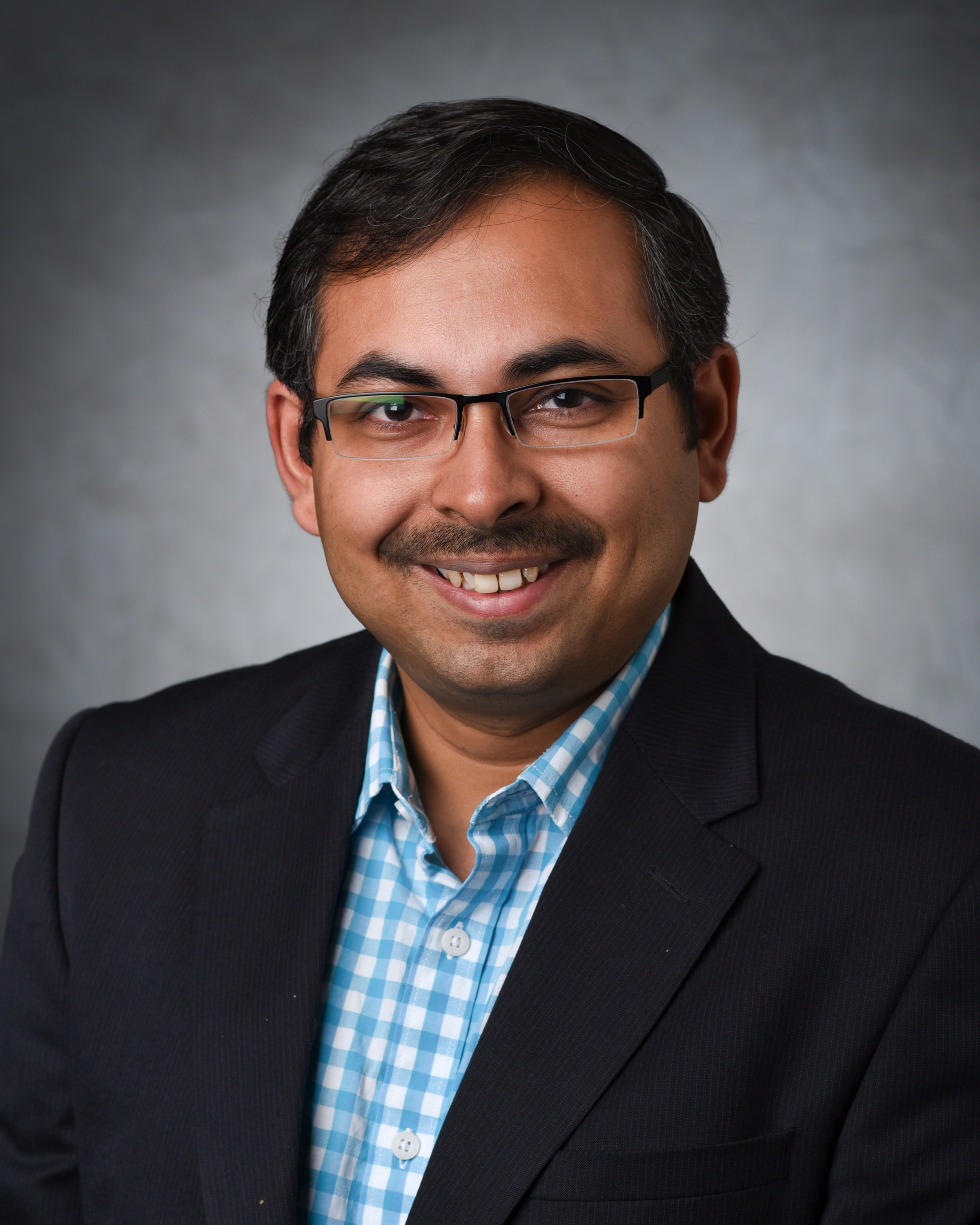}}]
{Nilanjan Ray Chaudhuri}(S’08-M’09-SM’16) received the Ph.D. degree in power systems from Imperial College London, London, UK in 2011. From 2005 to 2007, he worked in General Electric (GE) John F. Welch Technology Center. He came back to GE and worked in GE Global Research Center, NY, USA as a Lead Engineer during 2011 to 2014. Presently, he is an Associate Professor with the School of Electrical Engineering and Computer Science at Penn State, University Park, PA. He was an Assistant Professor with North Dakota State University, Fargo, ND, USA during 2014-2016. He is a member of the IEEE and IEEE PES. Dr. Ray Chaudhuri is the lead author of the book Multi-terminal Direct Current Grids: Modeling, Analysis, and Control (Wiley/IEEE Press, 2014). He served as an Associate Editor of the IEEE TRANSACTIONS ON POWER DELIVERY (2013 – 2019) and IEEE PES LETTERS (2016 - present). Dr. Ray Chaudhuri was the recipient of the National Science Foundation Early Faculty CAREER Award in 2016 and Joel and Ruth Spira Excellence in Teaching Award in 2019.
\end{IEEEbiography}

\end{document}